\documentclass[11pt]{article}
\pdfoutput=1
\usepackage{graphicx}
\usepackage{amssymb}
\usepackage{amsmath}
\usepackage[left]{lineno}
\usepackage{subcaption} 
\usepackage{booktabs}
\usepackage{multirow}
\usepackage[title]{appendix}
\usepackage{lineno}
\usepackage{relsize}
\usepackage{physics} 
\usepackage{siunitx}  
\usepackage{enumitem}
\setlist[itemize]{itemsep=0pt,topsep=\baselineskip} 
\usepackage{pgfplots}
\usepackage{pgfplotstable}
\usepackage{tikz,pgfplots}
\usepackage{hyperref}
\usepackage{xcolor}
\pgfplotsset{compat=1.14}
\newsavebox{\foobox}
\newcommand{\slantbox}[2][0]{\mbox{%
  \sbox{\foobox}{#2}%
  \hskip\wd\foobox
  \pdfsave
  \pdfsetmatrix{1 0 #1 1}%
  \llap{\usebox{\foobox}}%
  \pdfrestore
}}
\newcommand\unslant[2][-.25]{\slantbox[#1]{$#2$}}
\newcommand{\micro}{\unslant\mu}

\newcommand\T{\rule{0pt}{2.3ex}}       
\newcommand\B{\rule[-1.2ex]{0pt}{0pt}} 

\usepackage{url}
\usepackage{jheppub} 
\begin{document}

\pagenumbering{arabic}
\centerline{\LARGE EUROPEAN ORGANIZATION FOR NUCLEAR RESEARCH}
\vspace{10mm}
\begin{flushright}
CERN-EP-2024-343 \\
16 December 2024 \\
-- \\
Revised version: \\
25 February 2025 \\
\end{flushright}

\begin{center}
\Large{
\bf Observation of the {\boldmath $K^{+}\rightarrow\pi^{+}\nu\bar{\nu}$} decay \\ 
and measurement of its branching ratio
\\
\vspace{10mm}
}
\begin{NoHyper}
\renewcommand{\thefootnote}{\fnsymbol{footnote}}
The NA62 Collaboration
\footnote[1]{
Corresponding authors: F.~Brizioli, R.~Fiorenza, J.~Swallow, \\
email: francesco.brizioli@cern.ch, renato.fiorenza@cern.ch, joel.christopher.swallow@cern.ch
}
\end{NoHyper}

\end{center}
\vspace{5mm}

\begin{abstract}

\centerline{\textbf{Abstract}}

\vspace{5mm}

A measurement of the $K^{+}\rightarrow\pi^{+}\nu\bar{\nu}$ decay by the NA62 experiment at the CERN SPS is presented, using data collected in 2021 and 2022.
This dataset was recorded, after modifications to the beamline and detectors, at a higher instantaneous beam intensity with respect to the 2016--2018 data taking.
Combining NA62 data collected in 2016--2022, a measurement of 
$\mathcal{B}(K^{+}\rightarrow\pi^{+}\nu\bar{\nu}) = \left( 13.0^{+ 3.3}_{- 3.0} \right)\times10^{-11}$
is reported.
With $51$ signal candidates observed and an expected background of $18^{+3}_{-2}$ events,
$\mathcal{B}(K^{+}\rightarrow\pi^{+}\nu\bar{\nu})$ becomes the smallest branching ratio measured with a signal significance above $5\,\sigma$.

\end{abstract}
\vspace{10mm}

\begin{center}
\em{Accepted for publication in JHEP}
\end{center}

\clearpage

\tableofcontents

\newpage
\clearpage

\section{Introduction}
\label{sec:Intro}

The $K^{+}\rightarrow\pi^{+}\nu\bar{\nu}$ decay is a golden mode for flavour physics because of a high precision Standard Model (SM) description and a high sensitivity to new physics beyond the Standard Model (BSM).
This decay is a flavour changing neutral current process, short-distance dominated, and proceeds at lowest order in the SM through electroweak box and penguin diagrams dominated by $t$-quark exchange. 
The decay is highly suppressed due to the GIM mechanism and the CKM suppression of the $t\rightarrow d$ quark transition.
Using tree-level measurements of the CKM matrix elements as external inputs, the SM branching ratio is predicted to be $\mathcal{B}(K^{+}\rightarrow\pi^{+}\nu\bar{\nu})=(8.4\pm1.0)\times10^{-11}$~\cite{Buras:2015qea}, while using only meson mixing processes to eliminate the strong dependence on $|V_{cb}|$, the predicted branching ratio is found to be $(8.60\pm0.42)\times10^{-11}$~\cite{Buras:2022wpw}. Using a full CKM parameter fit, a value of $(7.86\pm0.61)\times10^{-11}$ is predicted~\cite{DAmbrosio:2022kvb}.
The precision is limited by the CKM parametric uncertainties coupled with the intrinsic theoretical uncertainty of approximately $3\%$.
The latter arises from QCD corrections to the top (charm) quark contribution at NLO (NNLO)~\cite{Buchalla:1998ba,Buras:2005gr}, NLO electroweak corrections~\cite{Brod:2010hi}, and the hadronic matrix element for the $K \rightarrow \pi$ transition extracted from $K^{+}\rightarrow\pi^{0}e^{+}\nu$ decay measurements~\cite{Brod:2010hi,Isidori:2005xm,Mescia:2007kn}.

The $K^{+}\rightarrow\pi^{+}\nu\bar{\nu}$ decay is sensitive to a variety of BSM effects, probing new physics at mass scales up to $\mathcal{O}(100\,\text{TeV})$~\cite{Buras:2015qea}.
Several BSM scenarios predict significant deviations of the branching ratio from the SM prediction, as well as correlations with other flavour observables and the corresponding decay mode of the neutral kaon $K_{L}\rightarrow\pi^{0}\nu\bar{\nu}$~\cite{Chen:2018ytc,Bobeth:2017ecx,Bobeth:2016llm,Endo:2016tnu,Endo:2017ums,Crivellin:2017gks,Blanke:2015wba,Bordone:2017lsy,Aebischer:2020mkv,Fajfer:2023nmz,Gorbahn:2023juq,Deppisch:2020oyx,Buras:2024ewl}.
A model-independent relationship between the branching ratios of the two decay modes is provided by the Grossman-Nir bound~\cite{Grossman:1997sk,PDG}:
$\mathcal{B}(K_{L}\rightarrow\pi^{0}\nu\bar{\nu}) \lesssim 4.3 \cdot \mathcal{B}(K^{+}\rightarrow\pi^{+}\nu\bar{\nu})$.
The direct upper limit of $\mathcal{B}(K_{L}\rightarrow\pi^{0}\nu\bar{\nu}) < 2.2 \times 10^{-9}$ at 90\% CL~\cite{KOTO:2024zbl}, set by the KOTO experiment, is two orders of magnitude above the SM predictions~\cite{Buras:2022wpw,DAmbrosio:2022kvb}.

The first $K^{+}\rightarrow\pi^{+}\nu\bar{\nu}$ branching ratio measurement was provided by the E787 and E949 experiments at BNL, using kaon decays at rest:
$\mathcal{B}(K^{+}\rightarrow\pi^{+}\nu\bar{\nu}) 
= (17.3^{+11.5}_{-10.5})\times10^{-11}$~\cite{BNL-E949:2009dza}.
The NA62 experiment at CERN was designed to study the $K^{+}\rightarrow\pi^{+}\nu\bar{\nu}$ decay with a decay-in-flight technique using a high-intensity secondary hadron beam.
The 2016--2018 dataset collected by NA62 produced the first evidence for this decay with a significance of $3.4\,\sigma$,  
measuring $\mathcal{B}(K^{+}\rightarrow\pi^{+}\nu\bar{\nu}) =
(10.6^{+4.1}_{-3.5})
\times10^{-11}$~\cite{PnnRun1Paper}. 
In the following, the branching ratio measurement using the data collected in 2021--2022 is presented as well as the combined result including 2016--2018 data.

\section{Beamline and detector}
\label{sec:Detector}

\begin{figure}
    \centering
    \includegraphics[width=1.0\linewidth]{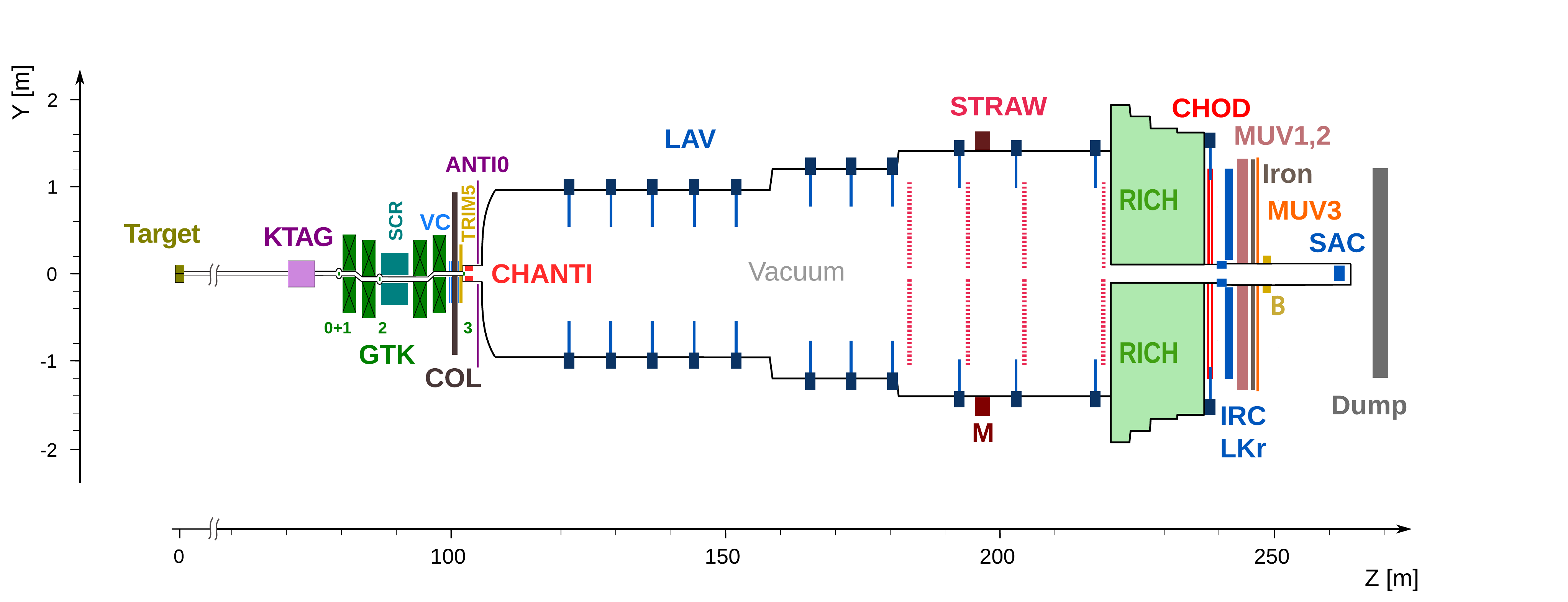}
    \vspace{-25pt}
    \caption{Schematic side view of the NA62 detector for data-taking from 2021 onwards.
    The MUV0 and HASC detectors are not visible in this view.
    }
    \label{fig:BealineAndDetectorSketch}
\end{figure}

A description of the NA62 beamline and detector is presented in~\cite{NA62DetectorPaper}. 
The upgraded setup used from 2021 onwards is displayed in figure~\ref{fig:BealineAndDetectorSketch}.
An unseparated secondary beam of $70\%$ $\pi^{+}$, $23\%$ protons and $6\%$ $K^{+}$ is created by directing $400\,\text{GeV}$ protons extracted from the CERN SPS onto a beryllium target in spills of 4.8 s
duration.
The target defines the origin of a right-handed coordinate system shown in figure~\ref{fig:BealineAndDetectorSketch}.
The beam central momentum is $75\,\text{GeV}/c$, with a momentum spread of $1\%$ (rms).
In the 2021--2022 dataset, each spill typically contains $3\times10^{12}$ protons.

Beam kaons are tagged by a differential Cherenkov counter (KTAG) with a $70\,\text{ps}$ time resolution. 
The KTAG used ${\rm N}_{2}$ gas at $1.75\,\text{bar}$ as a radiator medium within a $5\,\text{m}$ long vessel, and Cherenkov photons are detected by 384 photomultiplier tubes (PMTs) grouped in eight sectors.
The three-momenta of beam particles are measured by the Gigatracker (GTK), a spectrometer formed from four silicon pixel detector stations (with $300 \times 300\,\micro\text{m}^{2}$ pixels arranged in matrices with dimensions of $27 \times 60\,\text{mm}^{2}$) and two pairs of dipole magnets forming an achromat.
The GTK provides track momentum, direction and time  resolutions of $0.15\,\text{GeV}/c$, 16~$\micro\text{rad}$, and $100\,\text{ps}$, respectively.
The GTK0 station was added prior to the 2021 data-taking to improve tracking performance and pileup rejection.
In addition, the GTK2 station was moved to be upstream of the scraper magnet (SCR) to reduce the background from beam interactions.
The GTK3 station is preceded by the final collimator (COL), $1.2\,\text{m}$ thick and made of steel with outer dimensions of $1.7\times1.8\,\text{m}^{2}$ and a central race-track shaped $46 \times 76\,\text{mm}^{2}$ bore containing the beam. 
The collimator absorbs hadrons which are produced in upstream $K^{+}$ decays and not contained in the beam pipe.
A veto counter (VC) was installed in 2021 to detect particles produced in upstream $K^{+}$ decays. 
The VC consists of three planes of $14\,\text{cm}$ wide and $1\,\text{cm}$ thick horizontal scintillator bars each read out by a PMT at either end. 
Two planes (VC1 and VC2, with a $25\,\text{mm} = 4.5X_{0}$ thick lead plate in between) are located immediately upstream of COL, while VC3 is immediately downstream.
There are $11$ scintillator bars in VC1 and VC2 ($3$ above and $8$ below the beam pipe) and $10$ scintillator bars in VC2 ($3$ above and $7$ below the beam pipe);  each bar is $4\,\text{cm}$ high except those adjacent to the beam pipe, which are $2\,\text{cm}$ high.

This arrangement enables basic particle identification: muons traverse all stations, charged hadrons are detected in VC1 and VC2, and photons are detected in VC2 after converting in the lead plate. 
Products of inelastic interactions in GTK3 are detected by six stations of plastic scintillator bars (CHANTI) read out by silicon photomultipliers (SiPMs).
The ANTI0 hodoscope, installed in 2021 and formed of scintillator tiles read out by SiPMs, detects charged particles from the upstream region outside of the CHANTI acceptance.
 
The beam is delivered into a vacuum tank evacuated to a pressure of $10^{-6}\,\text{mbar}$, containing a fiducial volume (FV) defined as the region $105<Z<170\,\text{m}$. 
The probability of beam $K^{+}$ decay in the FV is $11.5\%$.
Charged particle momenta are measured with a resolution $\sigma_{p}/p = ( 0.30\oplus 0.005 \cdot p)\%$, with the momentum $p$ expressed in GeV/$c$,
by a magnetic spectrometer (STRAW) consisting of four straw chambers and a \SI{0.9}{\tesla\m} dipole magnet (M), which bends positively charged particles towards $X<0$. 
A $17\,\text{m}$ long ring-imaging Cherenkov detector (RICH), filled with neon gas at atmospheric pressure and read out with two arrays of PMTs, 
measures charged particle times with a typical resolution of $70\,\text{ps}$, 
provides the trigger reference time and is used for particle identification. 
Two scintillator hodoscopes (labelled CHOD in figure~\ref{fig:BealineAndDetectorSketch}) comprising a matrix of tiles (CHOD) and two planes of slabs (NA48-CHOD), both arranged in four quadrants, provide trigger signals and time measurements with $1\,\text{ns}$ and $200\,\text{ps}$ precision, respectively.

A set of photon veto detectors is designed to provide hermetic coverage of polar angles up to $50\,\text{mrad}$ from the beam axis, for photons emitted, for instance, in the decay chain $K^{+}\rightarrow\pi^{+}\pi^{0}$, $\pi^{0}\rightarrow\gamma\gamma$. Twelve large-angle veto (LAV) stations, ring-shaped electromagnetic calorimeters made of lead-glass blocks read out with PMTs, are arranged from $Z=121$ to $238\,\text{m}$ to detect photons emitted at $8.5$--$50\,\text{mrad}$. 
A 27 radiation-length thick, quasi-homogeneous liquid krypton
electromagnetic calorimeter (LKr) detects photons emitted at angles from $1$ to $8.5\,\text{mrad}$. 
The LKr is also used for particle identification, with energy resolution $\sigma_{E}/E = (4.8/\sqrt{E} \oplus 11/E \oplus 0.9)\%$ with $E$ expressed in GeV, spatial resolution of $1\,\text{mm}$, and time resolution between $0.5$ and $1\,\text{ns}$, depending on the energy deposited.
The intermediate-ring (IRC) and small-angle (SAC) lead/scintillator shashlik calorimeters are designed to detect photons emitted down to zero degrees in the forward direction. 
The IRC is located in front of the LKr, covering an annular region between $65$ and $135\,\text{mm}$ from the $Z$ axis. 
The SAC is located on-axis after a dipole magnet (B) which bends undecayed beam particles towards $X<0$ and into a beam dump.

Additional pion/muon discrimination capability is provided by two hadronic sampling calorimeters (MUV1,2) made from alternating layers of iron plates and scintillator strips, and an array of scintillator tiles (MUV3) with $400\,\text{ps}$ time resolution, located behind an $80\,\text{cm}$ iron wall.
Other veto detectors (MUV0 and HASC, located immediately downstream of the RICH and upstream of the SAC, respectively) provide additional background rejection.
A HASC station located at $X<0$ was augmented in 2021 by a second station at $X>0$ to enhance rejection capabilities.

\section{Data sample and trigger}
\label{sec:Dataset}

The 2016, 2017 and 2018 data were collected with mean instantaneous intensities of 240, 330 and 400~MHz, respectively, as measured by counting out-of-time GTK signals. 
The 2021 and 2022 data, used for this analysis, were collected at the design mean intensity of 580~MHz.
Data collected during the first second of the spill in 2021 are removed because of the systematic presence of spikes in instantaneous intensity, up to $10~\text{GHz}$.
In 2022, following improvements to the beam delivery systems, good quality data were collected consistently throughout the spills.

A two-stage trigger system is employed with successive hardware (L0) and software (L1) levels.  
Three trigger lines are used:
\begin{itemize}
    \item Minimum Bias (MB), to collect control samples of $K^{+}\rightarrow\mu^{+}\nu$ decays.
    \item Normalisation (NORM), to collect a normalisation sample of $K^{+}\rightarrow\pi^{+}\pi^{0}$ decays.
    \item Signal (PNN), to collect signal $K^{+}\rightarrow\pi^{+}\nu\bar{\nu}$ candidates.
\end{itemize}
\begin{table} 
    \centering
    \caption{Summary of trigger lines and trigger conditions.}
        \vspace{8pt}
    \resizebox{0.99\textwidth}{!}{ 
    \begin{tabular}{|r|l|c|c|c|} 
    \hline 
    Conditions & Requirements & MB & NORM & PNN \\
    \hline
    \multicolumn{5}{c}{L0} \T\B  \\
    \hline
    \texttt{RICH} & at least two signals in RICH & \checkmark & \checkmark & \checkmark \T\B \\
    \texttt{Q1} & at least one signal in the CHOD         & \checkmark & \checkmark & \checkmark \T\B \\
    $\overline{\text{ \texttt{MUV3}} }$ & no signals in MUV3 &  & \checkmark & \checkmark \T\B \\
    \texttt{UTMC} & fewer than five signals in the CHOD & & & \checkmark \T\B \\
    $\overline{\text{ \texttt{QX}} }$ & no signals in diagonally-opposite CHOD quadrants & & & \checkmark \T\B \\
    $\overline{\text{ \texttt{LKr40}} }$ & $E_{\text{LKr}}^{\rm total}<40\,\text{GeV}$ and fewer than two LKr clusters with $E>5\,\text{GeV}$ & & & \checkmark \T\B \\
    \hline
    \multicolumn{5}{c}{L1} \T\B  \\
    \hline
    \texttt{KTAG} & signals in at least five KTAG sectors & & \checkmark & \checkmark \T\B \\
    \texttt{STRAW-1TRK} & at least one positively charged STRAW track with $p<65\,\text{GeV}/c$ & & \checkmark & \checkmark \T\B \\
    \texttt{STRAW} & isolated $p<65\,\text{GeV}/c$ STRAW track in geometric acceptance & & & \checkmark \T\B \\
    $\overline{\text{ \texttt{LAV}} }$ & fewer than two signals in LAV2--11 & & & \checkmark \T\B \\
    \hline
    \end{tabular}
    }
    \label{tab:TriggerDefinitions}
\end{table}
A summary of the trigger conditions is given in table~\ref{tab:TriggerDefinitions} and further details can be found in~\cite{NA62TriggerPaper}.
The RICH provides a reference time; coincidences are required to be within $6.3\,\text{ns}$ of this reference time.
Downscaling factors of $D_\text{NORM}$ and $D_\text{MB}$ are applied at L0 to the NORM and MB trigger lines, respectively.

Monte Carlo simulations of particle interactions with the detector and its response are performed using a software package based on the GEANT4 toolkit~\cite{Geant4}.

\section{Selection}
\label{sec:Selection}

The signal selection for $K^{+}\rightarrow\pi^{+}\nu\bar{\nu}$ decays is applied to the sample collected by the PNN trigger line, while the selection for $K^{+}\rightarrow\pi^{+}\pi^{0}$ decays, used for normalisation, is applied to the sample collected by the NORM trigger line.

\subsection{Common selection criteria}
\label{sec:NormSelection}

A candidate beam $K^{+}$ is tagged by the KTAG and its momentum is measured by the GTK.
There must be coincident signals in at least five KTAG sectors, and the GTK momentum measurement must be in the range $72.7$--$77.2\,\text{GeV}/c$.

\begin{figure}[tb]
\centering
\includegraphics[width=0.49\linewidth]{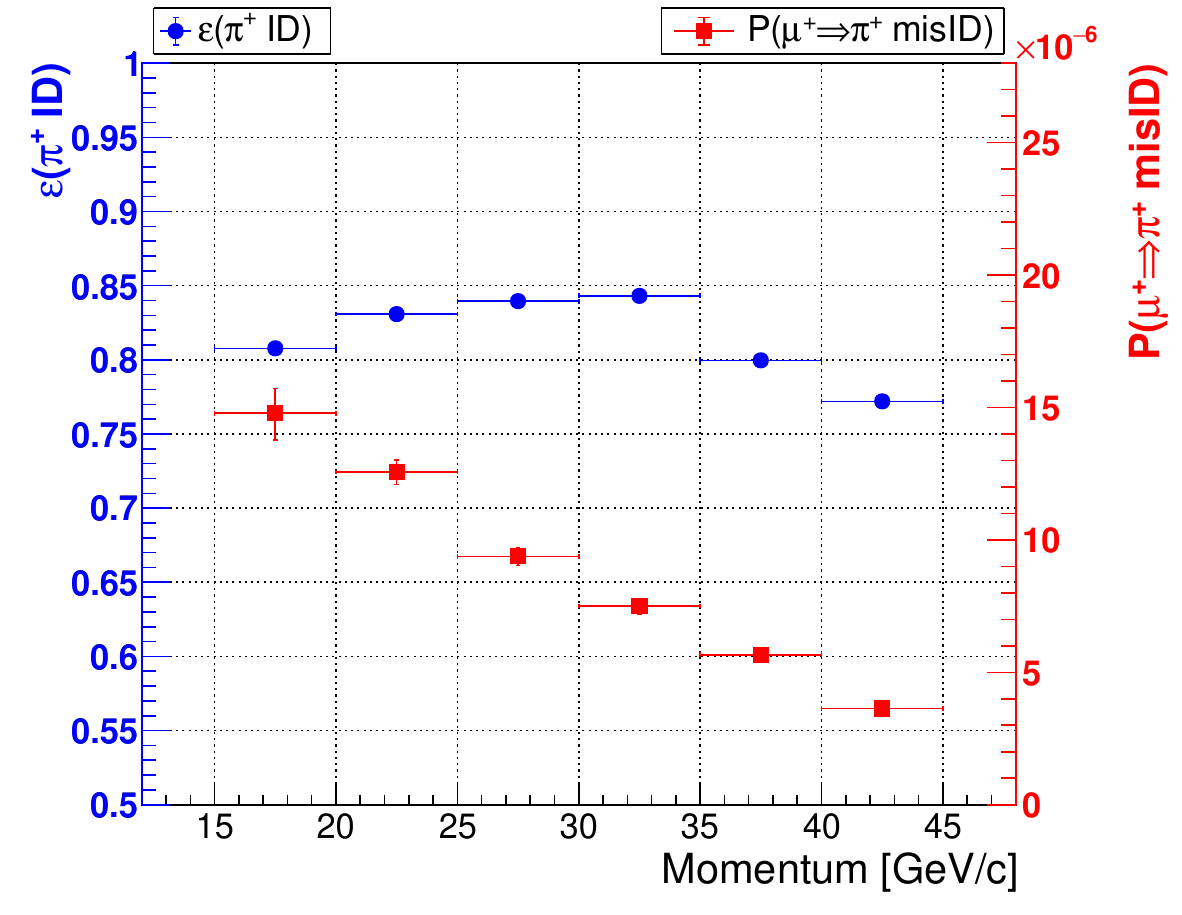}
\includegraphics[width=0.49\linewidth]{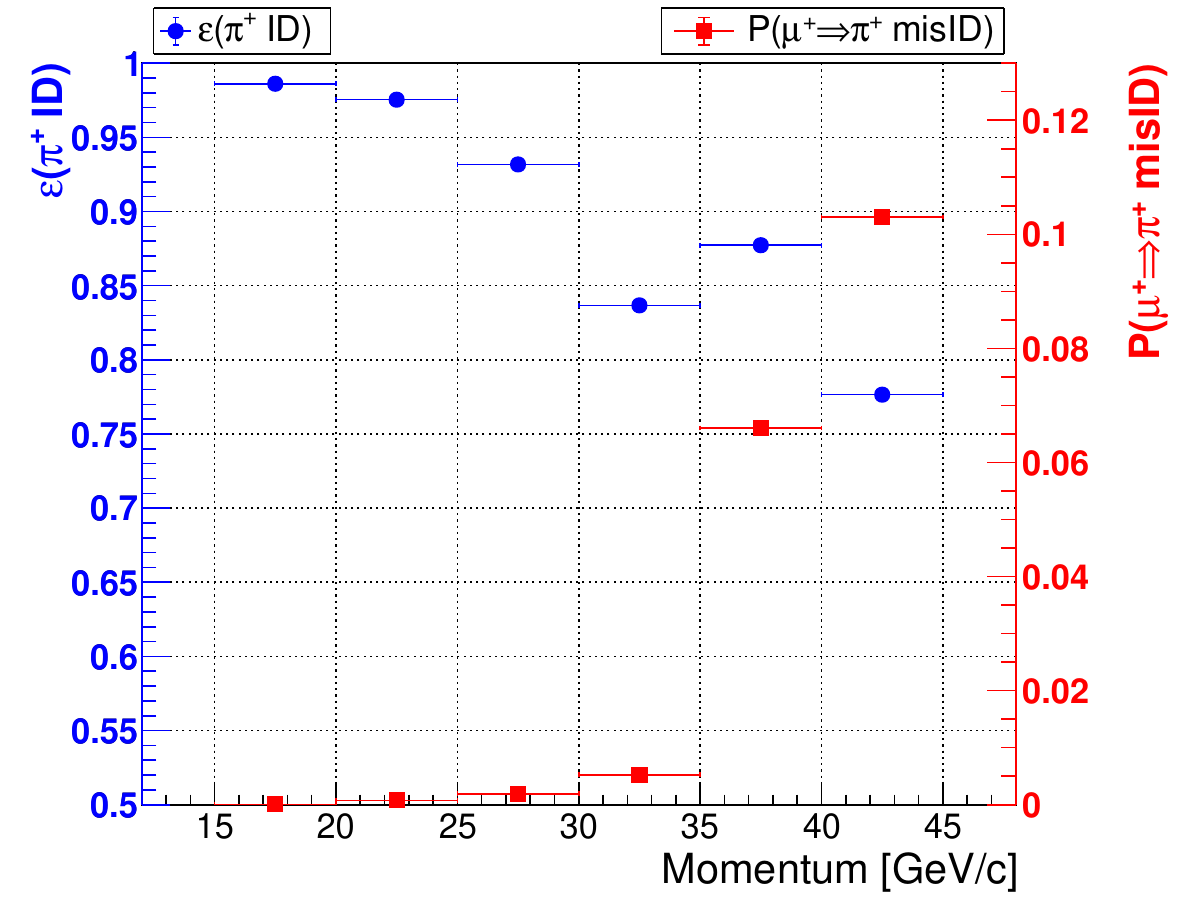}
\vspace{-5mm}
\caption{
Particle identification performance as a function of momentum, using information from LKr, MUV1,2 and MUV3 (left), and from RICH (right).
The $\pi^{+}$ identification efficiency is shown as blue circles (left vertical axis) and the probability of misidentifiaction of a $\mu^{+}$ as a $\pi^{+}$ is shown as red squares (right vertical axis).
}
\label{fig:PIDPerformance}
\end{figure}

The momentum of a $\pi^{+}$ candidate, $p_{\pi^{+}}$, measured by the STRAW is required to be in the range $15$--$45\,\text{GeV}/c$.  
The $\pi^{+}$ candidate is identified using the RICH, LKr, MUV1,2 and MUV3 information; 
the RICH additionally provides a time measurement with a $100\,\text{ps}$ time resolution.
A boosted decision tree (BDT) classifier uses calorimeter (LKr, MUV1,2) information for particle identification (PID) information.
No signals in MUV3 should be associated to the $\pi^{+}$ candidate.
The PID performance, quantified by the $\pi^{+}$ identification efficiency $\varepsilon(\pi^{+}\,\text{ID})$ and the probability of misidentifying a $\mu^{+}$ as a $\pi^{+}$, $\operatorname{P}(\mu^{+}\Rightarrow\pi^{+}\,\text{misID})$, is shown in figure~\ref{fig:PIDPerformance} as a function of momentum. 
The PID criteria are optimised in each momentum bin, 
leading to a non-monotonic variation.
Signals in the CHOD and NA48-CHOD associated with the $\pi^{+}$ candidate provide an additional time measurement.

The candidate $\pi^{+}$ is matched to a candidate beam $K^{+}$ using spatial and time information. 
By extrapolating the $K^{+}$ and $\pi^{+}$ trajectories from the GTK and STRAW, a vertex is defined as the mid-point of the closest distance of approach (CDA) segment between the two extrapolated tracks. 
The CDA must be less than $4\,\text{mm}$ and the longitudinal position of the vertex must be inside the FV. 
The KTAG, GTK and RICH times must all agree within $500\,\text{ps}$, while the NA48-CHOD and GTK times must agree within $1.1\,\text{ns}$.
Finally, a Bayesian discriminant is used to match the candidate $\pi^{+}$ to the $K^{+}$.
This discriminant uses: a probability density function of the CDA; a probability density function of $\Delta T_{\text{match}}$, proportional to the difference between the time of the GTK track and the average of the KTAG and RICH times;
a prior based on $N_{\text{GTK}}$, the number of GTK tracks within $3\,\text{ns}$ of the average of the RICH and KTAG times.
If $N_{\text{GTK}}>1$, the likelihood values are used to select the best match and reject events with overlapping GTK tracks.
The squared missing mass is evaluated as $m_{\rm miss}^{2} = (P_{K}-P_{\pi})^{2}$,  where $P_{K}$ and $P_{\pi}$ are the 4-momenta of the $K^{+}$ and $\pi^{+}$ candidates.
No in-time STRAW tracks forming a vertex with the $\pi^{+}$ candidate are allowed.

A set of veto conditions is applied against interactions and decays upstream of the FV. 
This includes rejecting GTK track segments and excluding events where $\pi^{+}$ candidates are consistent with originating from inside the beam pipe at GTK3. 
No VC signals should be present within $2\,\text{ns}$ of the $K^{+}$ time, except if a muon-like signature is observed with coincident signals in all three VC stations. 
No CHANTI signals should be present within $3\,\text{ns}$ of the $K^{+}$ or $\pi^{+}$ times. 
No ANTI0 signals geometrically compatible with the extrapolated $\pi^{+}$ candidate position should be present within $3\,\text{ns}$ of the $K^{+}$ time.
In addition, a condition is applied based on a BDT classifier, which uses spatial information from the $K^{+}$ and $\pi^{+}$ candidates, trained against interactions and decays upstream of the FV.

A set of kinematic and calorimetric conditions to veto $K^+ \rightarrow \mu^+ \nu \gamma$ events is described in section~\ref{sec:BackgroundRadiative}.

\subsection{Specific signal selection conditions}
\label{sec:VetosInSignalSel}

Photon veto criteria are applied to select signal candidates as follows:
no signals are allowed in the LAV stations downstream of the vertex within $3\,\text{ns}$ of the $\pi^{+}$ time; 
no signals from the IRC or SAC are allowed within $5\,\text{ns}$ of the $\pi^{+}$ time;
no energy clusters in the LKr are allowed at a distance exceeding $100 \,\text{mm}$ from the $\pi^{+}$ candidate impact point within an energy-dependent time window (with a width varying between $5$ and $10\,\text{ns}$) of the $\pi^{+}$ time.

Multiplicity veto criteria are applied as follows:
no additional coincident signals (not associated to the $\pi^{+}$ candidate) are allowed in any two of the CHOD, NA48-CHOD and LKr detectors; 
no MUV0 or HASC signals coincident with the $\pi^{+}$ candidate are allowed;
no additional STRAW track segments are allowed which form a vertex with the $\pi^{+}$ candidate.

\begin{figure}[tb]
\centering
\includegraphics[width=0.49\linewidth]{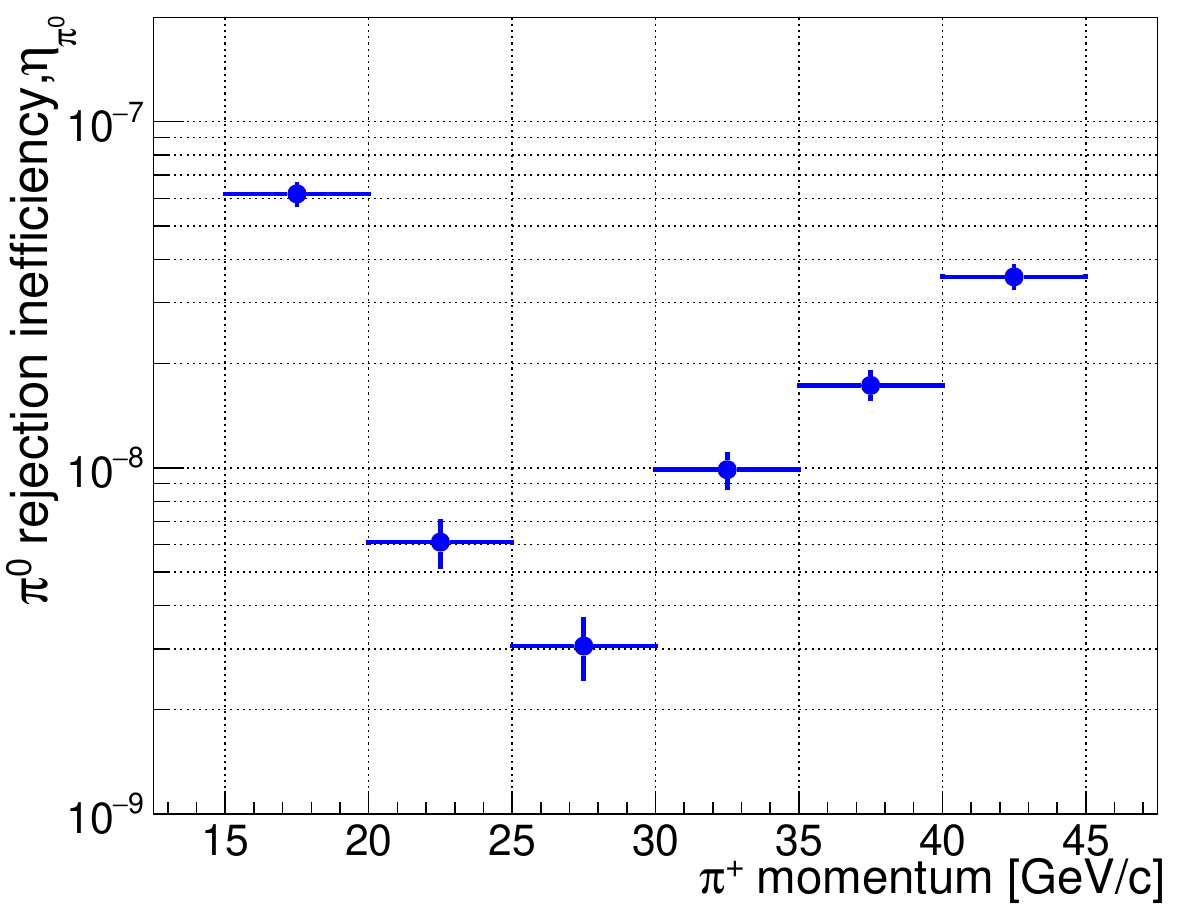} 
\includegraphics[width=0.49\linewidth]{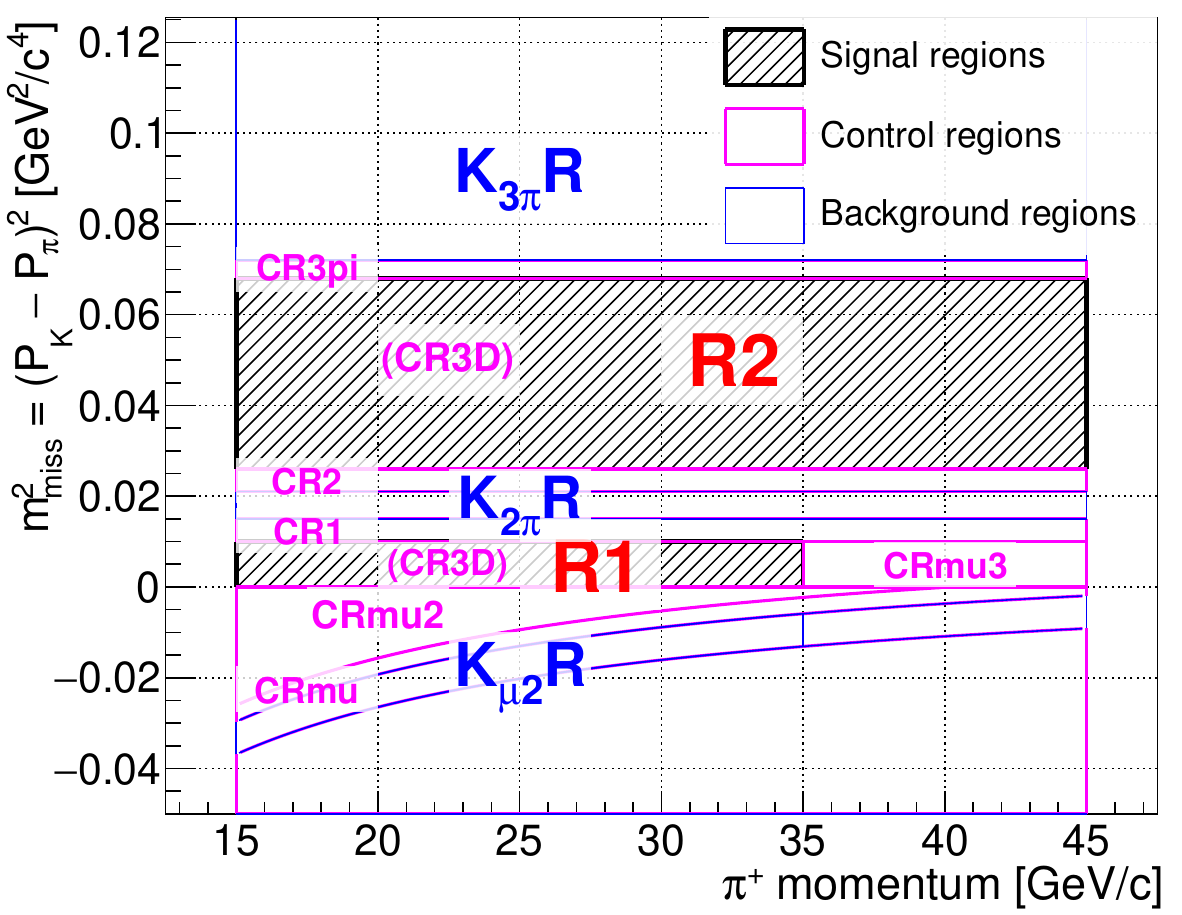} 
\caption{
Left: $\pi^{0}$ rejection inefficiency as a function of the $\pi^+$ momentum.
Right: definitions of kinematic regions in the $(p_{\pi^{+}},m_{\rm miss}^{2})$ plane. 
Region CR3D is the same as the signal region in this projection, but contains events outside the 3-dimensional signal regions definition.
}
\label{fig:pi0Rejection}
\end{figure}

The $\pi^{0}$ rejection inefficiency, defined as the probability that a normalisation $K^{+}\rightarrow\pi^{+}\pi^{0}$ decay is not rejected by the photon or multiplicity veto criteria, depends on the $\pi^+$ momentum, as shown in figure~\ref{fig:pi0Rejection}-left; on average it is measured to be $\eta_{\pi^{0}} = (1.72\pm0.07)\times10^{-8}$.

\subsection{Kinematic regions}
Normalisation $K^{+}\rightarrow\pi^{+}\pi^{0}$ candidates are selected in the $m_{\rm miss}^{2}$ range $0.010$--$0.026\,\text{GeV}^{2}/c^{4}$, centred at the $\pi^0$ mass squared~\cite{PDG}, which has a resolution of  $0.001\,\text{GeV}^{2}/c^{4}$~\cite{PnnRun1Paper}.

The definitions of the kinematic regions~\cite{Pnn2017paper,PnnRun1Paper} are shown in figure~\ref{fig:pi0Rejection}-right. 
In the signal selection, two \textit{signal regions} are defined: R1 with $m_{\rm miss}^{2} $ of $0.000$--$0.010\,\text{GeV}^{2}/c^{4}$ and $p_{\pi^{+}} $ of $15$--$35\,\text{GeV}/c$; R2 with $m_{\rm miss}^{2} $ of $0.026$--$0.068\,\text{GeV}^{2}/c^{4}$ and $p_{\pi^{+}} $ of $15$--$45\,\text{GeV}/c$. 
The definition of the signal regions includes additional constraints based on squared missing mass observables calculated analogously to $m_{\rm miss}^{2}$ after replacing: the GTK $K^{+}$ momentum with the average beam momentum;  
the STRAW $\pi^{+}$ momentum with the RICH $\pi^{+}$ momentum (using the RICH as a velocity spectrometer and the $\pi^{+}$ direction measured by the STRAW).

A set of \textit{control regions} is established (CR1, CR2, CRmu, CRmu2, CRmu3, CR3pi, CR3D), located between the signal regions and three \textit{background regions}, $K_{\mu2}R$, $K_{2\pi}R$ and $K_{3\pi}R$, which contain $K^{+}\rightarrow\mu^{+}\nu$, $K^{+}\rightarrow\pi^{+}\pi^{0}$ and $K^{+}\rightarrow\pi^{+}\pi^{+}\pi^{-}$ decays, respectively. 
The control regions are used to validate the background estimates.

\section{Signal sensitivity}
\label{sec:SignalSensitivity}

The effective number of $K^{+}$ decays is evaluated as
\begin{equation}
    N_{K} = \frac{(1-\varepsilon_{B}) \sum_{i} N_{\pi\pi}^{i}\,D_\text{NORM}^{i}}{\mathcal{B}_{\pi\pi}\,A_{\pi\pi}} = \frac{N_{\pi\pi}^{\rm eff} \cdot 400}{\mathcal{B}_{\pi\pi}\,A_{\pi\pi}}\,,
\end{equation}
where 
$N_{\pi\pi}^{i}$ is the number of normalisation events selected in spill $i$, 
$D_{\rm NORM}^{i}$ is the downscaling of the NORM trigger line (typically 400),
$\mathcal{B}_{\pi\pi}$ is the branching ratio of the normalisation $K^{+}\rightarrow\pi^{+}\pi^{0}, \, \pi^0\rightarrow\gamma\gamma$ decay chain~\cite{PDG}, and
$A_{\pi\pi}$ is the normalisation selection acceptance.
The quantity $\varepsilon_{B}$ is the background contamination of the normalisation sample, dominated by the $K^{+}\to\pi^{+}\pi^{0}$, $\pi^{0}\to e^{+}e^{-}\gamma$ decay chain, and estimated with simulations to be $(0.2\pm0.2)\%$.

The single event sensitivity is the $K^{+}\rightarrow\pi^{+}\nu\bar{\nu}$ branching ratio which would lead to an expectation of a single signal event, and is evaluated as
\begin{equation}
    \mathcal{B}_\text{SES} = \frac{1}{N_{K}\,A_{\pi\nu\bar{\nu}}\, \varepsilon_\text{trig}\,\varepsilon_\text{RV}}\,.
     \label{eqn:SES}
\end{equation}
Here $A_{\pi\nu\bar{\nu}}$ is the signal selection acceptance;
$\varepsilon_\text{trig} = \varepsilon_\text{PNN} / \varepsilon_\text{NORM}$ is the trigger efficiency ratio of the PNN and NORM trigger lines for the signal and normalisation samples, respectively; 
and $1 - \varepsilon_\text{RV}$ is the probability that a signal event is rejected by the veto conditions due to the presence of unrelated activity ($\varepsilon_\text{RV}$ is referred to as random veto efficiency).

The analysis is performed in six $5\,\text{GeV}/c$ wide bins of $\pi^+$ momentum in the range 15--45~GeV/$c$.
The single event sensitivity is converted to the number of expected SM $K^{+}\rightarrow\pi^{+}\nu\bar{\nu}$ events, assuming a given value for the SM branching ratio $\mathcal{B}_{\pi\nu\bar{\nu}}^{\text{SM}}$, in a momentum bin $p_{i}$, as
\begin{equation}
    N_{\pi\nu\bar{\nu}}^{\text{SM}}(p_{i}) = \frac{\mathcal{B}_{\pi\nu\bar{\nu}}^{\text{SM}}}{\mathcal{B}_\text{SES}(p_{i})} = 
    \frac{\mathcal{B}_{\pi\nu\bar{\nu}}^{\text{SM}}}{\mathcal{B}_{\pi\pi}}
   \frac{ N_{\pi\pi}^{\rm eff}(p_{i})\cdot 400}{A_{\pi\pi}(p_{i})}
    A_{\pi\nu\bar{\nu}}(p_{i})\,
    \varepsilon_\text{trig}(p_{i})\,
    \varepsilon_\text{RV}\,.
    \label{eqn:Npnn}
\end{equation}
The evaluation of each factor is described in the following and results, summed or averaged over momentum bins, are summarised in table~\ref{tab:SESSummaryTable}. The precision of the $\mathcal{B}_\text{SES}$ estimation is $3.5\%$, which is a  significant improvement with respect to the 2018 analysis.
The improvement is due to the higher precision of the trigger efficiency and the better $\varepsilon_\text{RV}$ evaluation strategy. 
The expected number of SM signal events per SPS spill in 2022 data is $2.5\times10^{-5}$, to be compared to $1.7\times10^{-5}$ in the 2018 data analysis~\cite{PnnRun1Paper}.

\begin{table} 
    \centering
    \caption{Signal sensitivity inputs for the total 2021--2022 data sample. 
    Using equation~\ref{eqn:Npnn}, $N_{\pi\nu\bar{\nu}}^{\text{SM}}$ is evaluated assuming $\mathcal{B}_{\pi\nu\bar{\nu}}^{\text{SM}} = 8.4\times10^{-11}$.}
        \vspace{8pt}
    \begin{tabular}{|cl|c|} 
    \hline
     & Factor & Value \T\B \\
    \hline
    \hline 
    $N_{\pi\pi}^{\rm eff}$ & Effective number of normalisation events & $(1.953\pm0.005)\times10^{8}$ \T\B \\ 
    $A_{\pi\pi}$ & Normalisation acceptance & $(13.410\pm0.005)\%$ \T\B \\
    $N_{K}$ & Effective number of $K^{+}$ decays & $(2.85\pm0.01)\times10^{12}$ \T\B \\ 
    \hline
    $A_{\pi\nu\bar\nu}$ & Signal acceptance & $(7.62\pm0.22)\%$ \T\B \\
    $\varepsilon_\text{trig}$ & Trigger efficiency ratio & $(85.9\pm1.4)\%$ \T\B \\
    $\varepsilon_\text{RV}$ & Random veto efficiency & $(63.2\pm0.6)\%$ \T\B \\ 
    \hline
    $\mathcal{B}_\text{SES}$ & Single event sensitivity & $(8.48\pm0.29)\times10^{-12}$ \T\B \\  
    $N_{\pi\nu\bar{\nu}}^{\text{SM}}$ & Number of expected SM $K^{+}\rightarrow\pi^{+}\nu\bar{\nu}$ events & $9.91\pm0.34$ \T\B \\
    \hline  
    \end{tabular} 
    \label{tab:SESSummaryTable}
\end{table}

\subsection{Acceptances}
\begin{figure}[tb]
\centering
\includegraphics[width=0.49\linewidth]{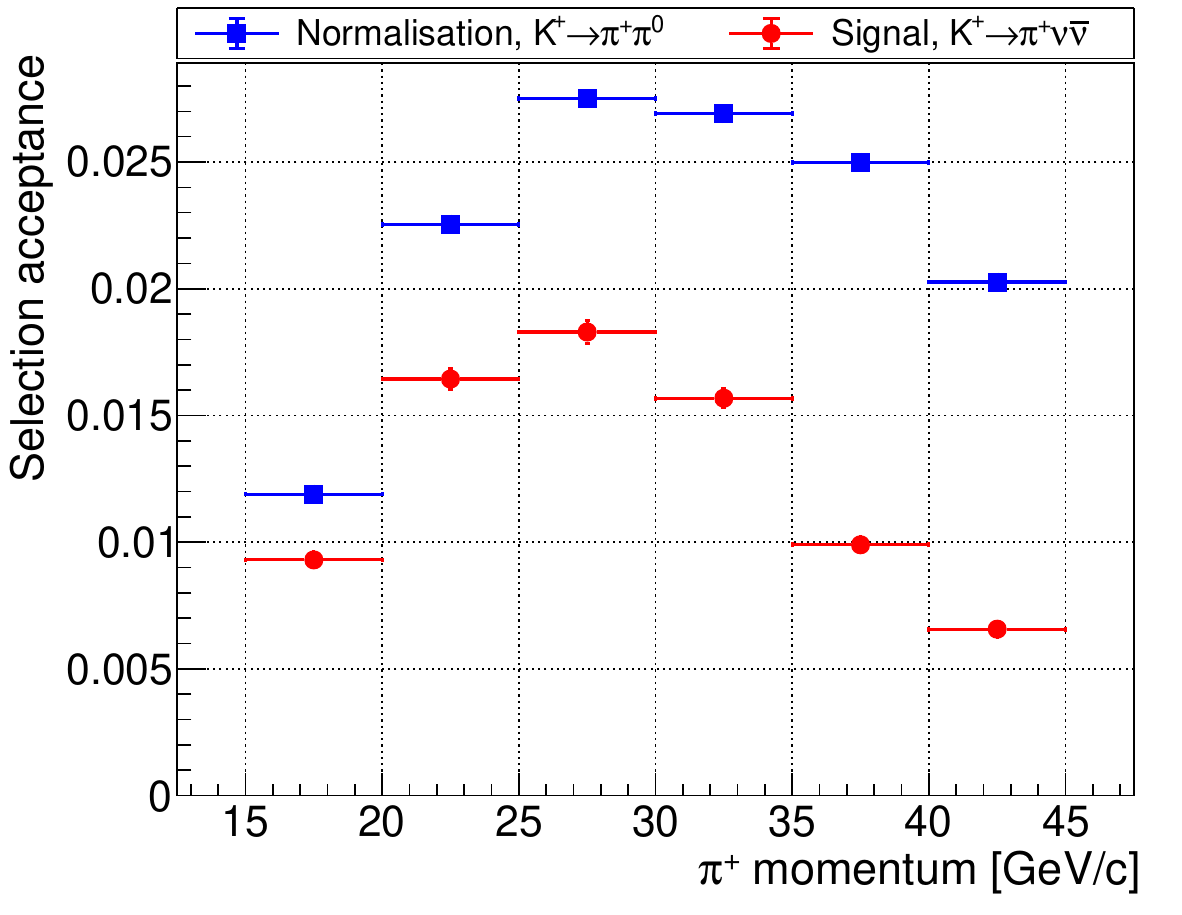}
\includegraphics[width=0.49\linewidth, trim=0 -10pt 0 0]{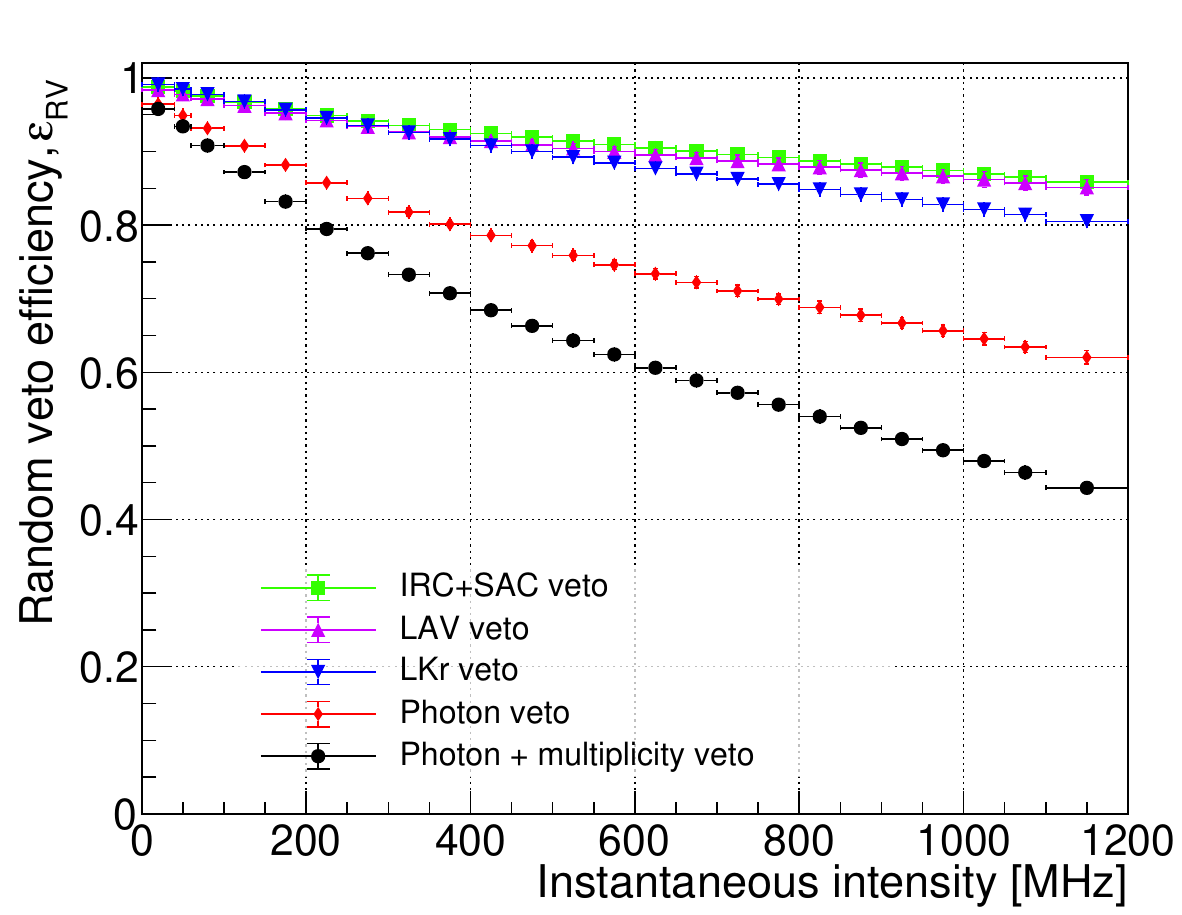} 
\caption{
Left: selection acceptances, $A_{\pi\pi}$ and $A_{\pi\nu\bar{\nu}}$, as functions of $\pi^{+}$ candidate momentum displayed as blue squares and red circles, respectively.
Right: random veto efficiency as a function of the instantaneous beam intensity.
}
\label{fig:SES}
\end{figure}

The acceptances are measured using simulated samples without pileup, of SM signal $K^{+}\rightarrow\pi^{+}\nu\bar{\nu}$ and normalisation $K^{+}\rightarrow\pi^{+}\pi^{0}, \, \pi^0\rightarrow\gamma\gamma$ decays.
Results are shown as a function of $\pi^{+}$ momentum in figure~\ref{fig:SES}-left. 
Systematic uncertainties in $A_{\pi\pi}$ and $A_{\pi\nu\bar{\nu}}$, associated with the simulation of $K^{+}$--$\pi^{+}$ matching and particle identification, cancel in the ratio in equation~\ref{eqn:Npnn}. 
A relative systematic uncertainty of $2.8\%$ is applied to $A_{\pi\nu\bar{\nu}}$ in each $\pi^{+}$ momentum bin, accounting for the effects that do not cancel in the acceptance ratio, mainly due to the photon and multiplicity veto conditions (see section~\ref{sec:VetosInSignalSel}).

\subsection{Random veto efficiency}
The random veto efficiency, $\varepsilon_\text{RV}$, accounts for that part of the selection specific to the signal, as the contribution from the normalisation selection cancels in the ratio. 
The additional activity, and hence $\varepsilon_\text{RV}$, is not associated with the $\pi^{+}$ track and depends only on the instantaneous beam intensity, and is therefore independent of $p_{\pi^{+}}$. 
A control sample of $K^{+}\rightarrow\mu^{+}\nu$ decays is used to measure $\varepsilon_\text{RV}$ as the fraction of these single-track events which are rejected due to the photon and multiplicity veto conditions described in section~\ref{sec:VetosInSignalSel}.  
To avoid bias due to additional signals in veto detectors from the $\mu^{+}$, the random veto efficiency is calculated as $\varepsilon_\text{RV} = \varepsilon_\text{RV}^\text{data}/\varepsilon_\text{RV}^\text{MC}$ where $\varepsilon_\text{RV}^\text{data}$ and $\varepsilon_\text{RV}^\text{MC}$ are measured in data and in a simulated sample of $K^{+}\rightarrow\mu^{+}\nu$ decays without pileup, respectively.
Results are shown in figure~\ref{fig:SES}-right as a function of instantaneous beam intensity.

\subsection{Trigger efficiencies}
\begin{figure}[tb]
\centering
\includegraphics[width=0.49\linewidth]{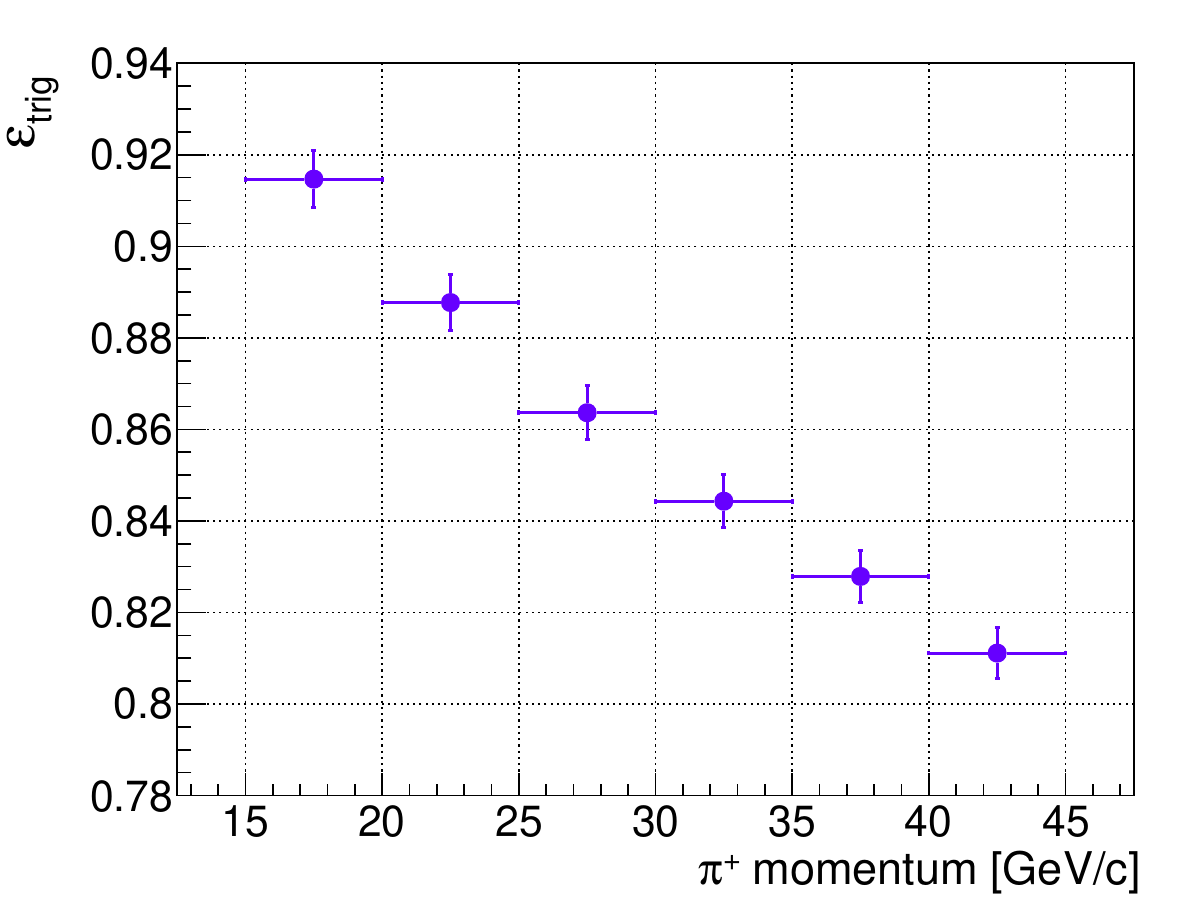}
\includegraphics[width=0.49\linewidth]{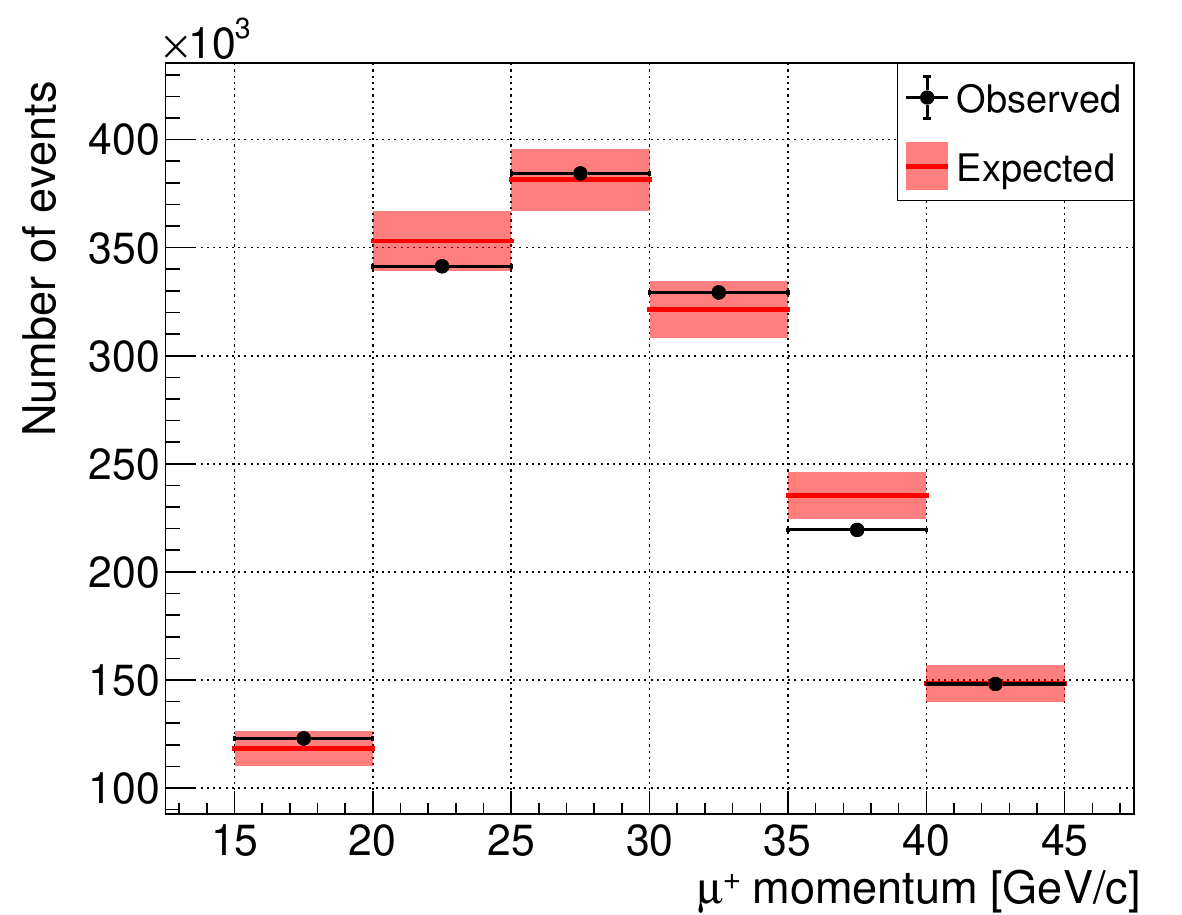}
\caption{
Left: trigger efficiency ratio $\varepsilon_\text{trig}$ as a function of the of $\pi^+$ momentum. 
Right: expected and observed numbers of $K^{+} \rightarrow \mu^{+}\nu$ events in trigger efficiency validation samples, as a function of the $\mu^+$ momentum.
}
\label{fig:SES_Trig}
\end{figure}

The efficiencies of common trigger conditions cancel in the ratio $\varepsilon_\text{trig}$.
The remaining inefficiency, mostly arising from the $\overline{\texttt{LKr40}}$ condition at L0 ($10\%$) and the $\overline{\texttt{LAV}}$ ($2.5\%$) condition at L1, is measured using the NORM trigger line, following the procedure described in~\cite{Pnn2017paper}. 
Results of the $\varepsilon_\text{trig}$ measurement are displayed as a function of momentum in figure~\ref{fig:SES_Trig}-left. 
The strong dependence of $\varepsilon_{\rm trig}$ on the $\pi^{+}$ candidate momentum arises from the $\overline{\texttt{LKr40}}$ condition, since at larger $p_{\pi^{+}}$ the probability to reach the $40\,\text{GeV}$ energy threshold is higher.
Conversely, the $\overline{\texttt{LAV}}$ inefficiency is primarily dependent on the instantaneous beam intensity.

The accuracy and precision of the trigger efficiency measurements are tested by comparing the expected and observed numbers of $K^{+}\rightarrow\mu^{+}\nu$ decays collected by the PNN trigger and satisfying the signal selection, without applying RICH PID criteria and selecting the $K^{+}\rightarrow\mu^{+}\nu$ background kinematic region (figure~\ref{fig:pi0Rejection}-right: $K_{\mu2}R$). The expected number of events is calculated as  
\begin{equation}
    N_{\mu\nu}^\text{PNN} = N_{\mu\nu}^\text{MB} D_\text{MB} \frac{\varepsilon_\text{PNN}}{\varepsilon_\text{MB}}\,\,.
    \label{eqn:NmunuTrigSyst}
\end{equation}
Here $N_{\mu\nu}^\text{MB}$ is the number of $K^{+}\rightarrow\mu^{+}\nu$ events collected by the MB trigger and satisfying the same selection criteria,  
$D_\text{MB}=600$ is the downscaling factor of the MB trigger, 
and the last term is the ratio of trigger efficiencies for the PNN and MB trigger lines. 
There are fewer common components between the MB and PNN trigger lines than between NORM and PNN, meaning the cancellation in this ratio is less significant.
Results of the validation are shown in figure~\ref{fig:SES_Trig}-right.
The expected and observed numbers of events are found to be consistent, given the uncertainties in the measurements. Therefore, no additional systematic uncertainty is applied.

\section{Background evaluation}
\label{sec:BackgroundStudies}

\def\generalregion{R} 

Background from $K^{+}$ decays in the FV arises if: (1) the $m_{\rm miss}^{2}$ is (mis)reconstructed to be inside the signal regions; (2) a charged track is misidentified as a $\pi^{+}$ or additional particles from the decay are not detected. 
In addition, there is an `upstream background' due to decays and interactions upstream of the FV with misreconstruction or mismatching of the candidate $\pi^{+}$ downstream.

\subsection{Background from main $K^+$ decay modes}
For the $K^{+}\rightarrow\mu^{+}\nu$, $K^{+}\rightarrow\pi^{+}\pi^{0}$ and $K^{+}\rightarrow\pi^{+}\pi^{+}\pi^{-}$ decays the numbers of expected background events are given by 
multiplying the number of events satisfying the full signal selection in the corresponding background region $\generalregion=K_{\mu2}R,K_{2\pi}R,K_{3\pi}R$ by the kinematic tail fraction, $f_\text{kin}$, defined as the ratio between the number of events in the signal region and the number of events in the background region $\generalregion$, evaluated in a dedicated control sample.

\begin{figure}[tb]
\centering
\includegraphics[width=0.49\linewidth]{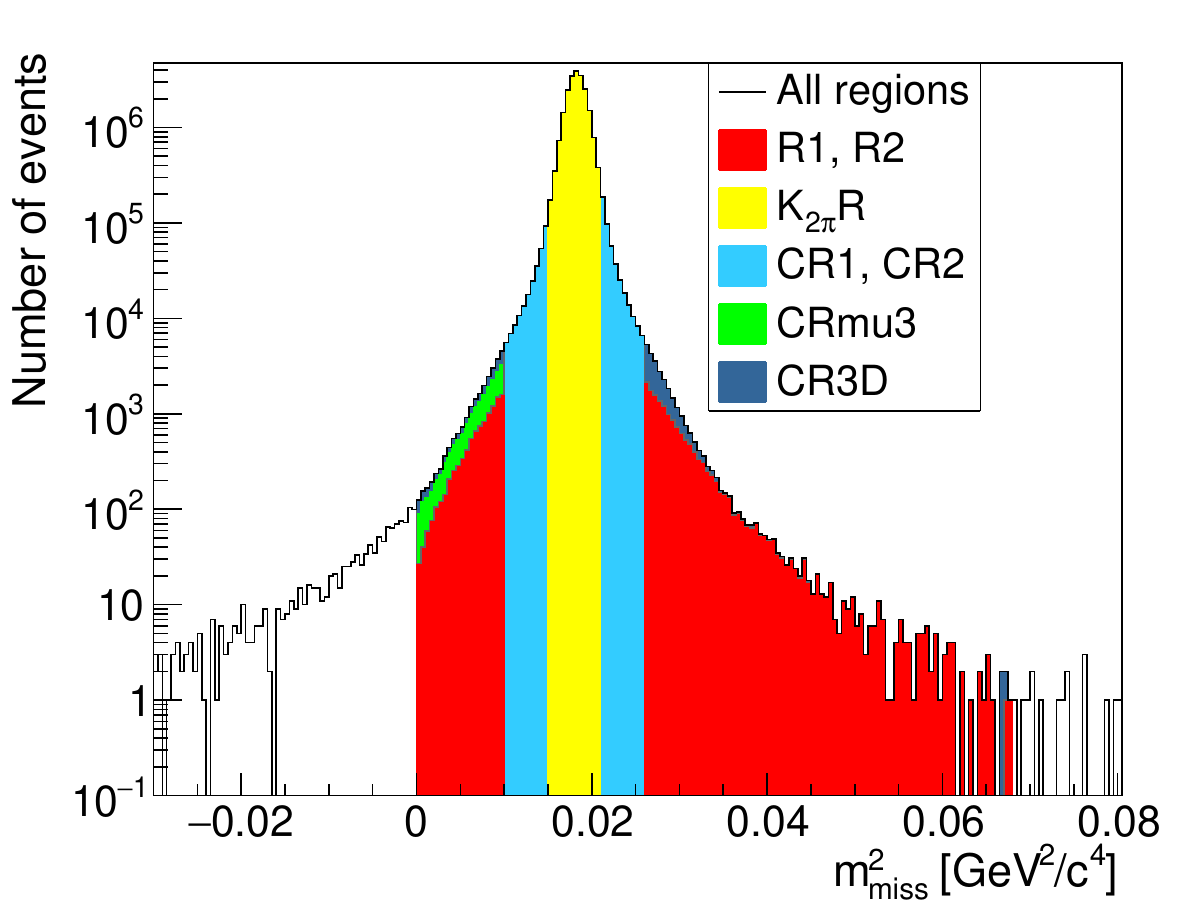}
\includegraphics[width=0.46\linewidth]{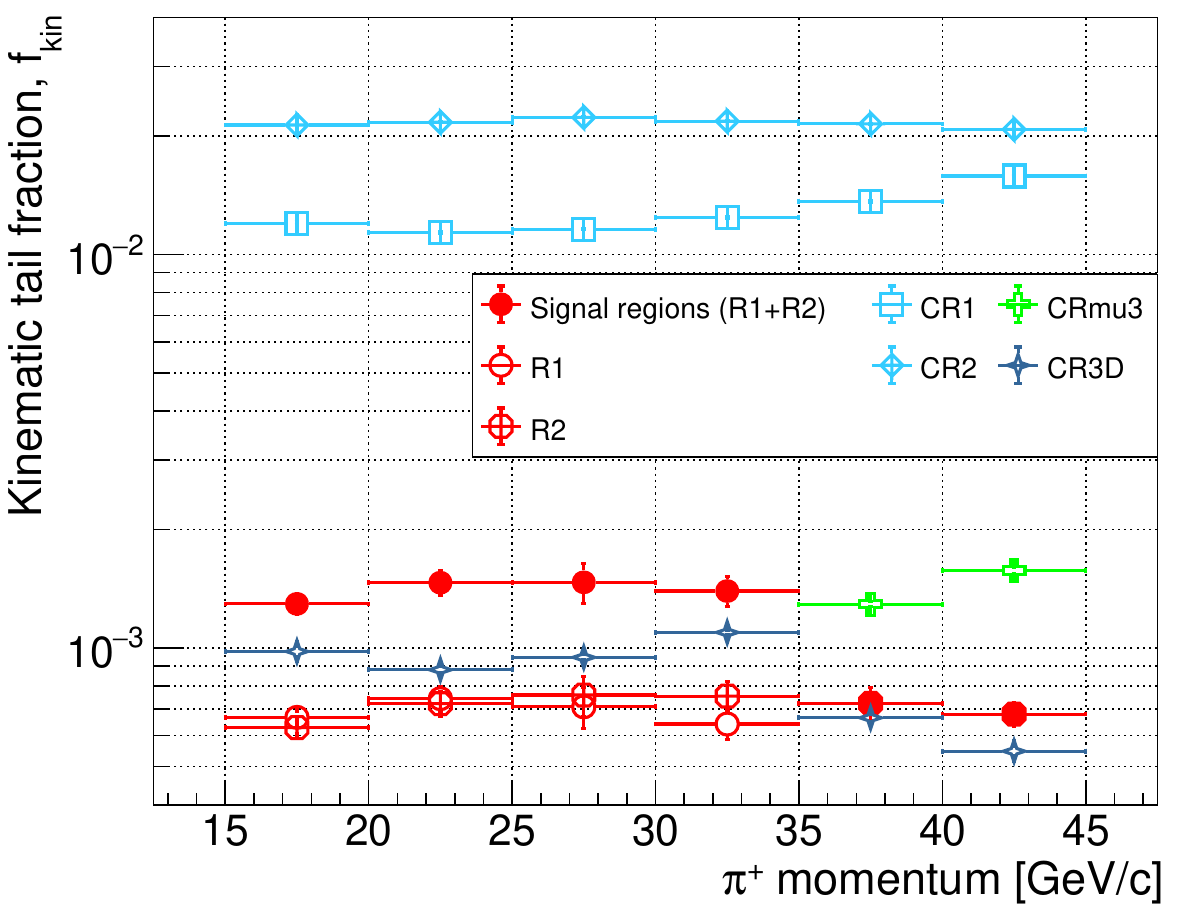}
\caption{
Left: distribution of $m_{\rm miss}^{2}$ of the kinematic tails data control sample for $K^{+}\rightarrow\pi^{+}\pi^{0}$ decays.
The colours show the contributions from the kinematic regions defined in figure~\ref{fig:pi0Rejection}-right. The average value of $f_\text{kin}(K^{+}\rightarrow\pi^{+}\pi^{0})$ is calculated as the ratio between the red area and the yellow area (and similarly for the control regions).
Right: $f_\text{kin}(K^{+}\rightarrow\pi^{+}\pi^{0})$ as a function of the $\pi^+$ momentum, evaluated for signal regions (red) and for control regions (by exchanging the signal region with the corresponding control region).
}
\label{fig:KinematicTailsSamples}
\end{figure}

For $K^{+}\rightarrow\pi^{+}\pi^{0}$ decays, the control sample is selected by 
reconstructing a $\pi^{0}\rightarrow\gamma\gamma$ decay by detecting exactly two photons in the LKr~\cite{Pnn2017paper}, 
and applying the conditions of the normalisation selection.
The $m_\text{miss}^{2}$ distribution of this control sample is shown in figure~\ref{fig:KinematicTailsSamples}-left.
The obtained values of $f_\text{kin}(K^{+}\rightarrow\pi^{+}\pi^{0})$ as functions of the $\pi^+$ momentum are shown in figure~\ref{fig:KinematicTailsSamples}-right.
The average value is $f_\text{kin}(K^{+}\rightarrow\pi^{+}\pi^{0}) = (1.20\pm0.01)\times10^{-3}$, and the total expected background is $N_{b}(K^{+}\rightarrow\pi^{+}\pi^{0}) = 0.76\pm0.04$.
Similarly, the background for each control region is estimated using the corresponding $f_{\text{kin}}$ values.

For $K^{+}\rightarrow\mu^{+}\nu$ decays, the control sample is selected by requiring calorimetric $\mu^{+}$ PID criteria are satisfied with a signal present in the MUV3 associated with the track.
However, the RICH $\pi^{+}$ PID conditions must also be satisfied to take into account the correlation with the signal region definition.
The average value is $f_\text{kin}(K^{+}\rightarrow\mu^{+}\nu) = (1.6\pm0.6)\times10^{-5}$, and the total expected background is $N_{b}(K^{+}\rightarrow\mu^{+}\nu) = 0.87\pm0.19$.
The background from the $K^{+}\rightarrow\mu^{+}\nu$, $\mu^{+}\rightarrow e^{+}\nu\bar{\nu}$ decay chain is found from simulations to be negligible.

For $K^{+}\rightarrow\pi^{+}\pi^{+}\pi^{-}$ decays, the kinematic tail fraction is evaluated using simulations: the average value is $f_\text{kin}(K^{+}\rightarrow\pi^{+}\pi^{+}\pi^{-}) = (6\pm2)\times10^{-6}$, and the total expected background is $N_{b}(K^{+}\rightarrow\pi^{+}\pi^{+}\pi^{-}) = 0.11\pm0.03$.

\label{sec:BackgroundRadiative}
The procedure described above does not fully include the contributions from the respective radiative decays, and therefore corrections are applied as described in the following.

Because of the veto on photons other than those from $\pi^{0}\to\gamma\gamma$ decays, the $K^{+}\rightarrow\pi^{+}\pi^{0}$ kinematic-tails control sample does not include events where a radiative photon is detected in LAV, LKr, IRC or SAC. Independent studies~\cite{Pnn2017paper} of single photon rejection capabilities demonstrate that such radiative decays have an additional factor $30$ rejection due to the extra photon. A simulation-driven estimation of the kinematic tail fraction of such radiative decays leads to an expected background of $0.07\pm0.01$.

Radiative $K^{+}\rightarrow\mu^{+}\nu\gamma$ decays are included in the $K^{+}\rightarrow\mu^{+}\nu$ control sample (where the same photon veto conditions as in the signal selection are applied) except for the specific case in which a high-momentum ($p \gtrsim 35\,\text{GeV}/c$) muon and a photon (with energy $E_{\gamma}\gtrsim5\,\text{GeV}$) overlap in the LKr forming a single energy cluster that may lead to a calorimetric misidentification as a $\pi^{+}$. 
This misidentification probability is found to be higher in the 2021--2022 data than in the previous 2016--2018 analysis, due to a performance degradation of the PID at higher intensities (particularly because of the higher occupancy in MUV1,2).
A significant excess of events in R2 in the 2021--2022 data, relative to the 2016--2018 data, was observed at $p_{\pi^{+}}>35\,\text{GeV}/c$. 
Data control samples and simulations were then used to identify and study this additional background to enable a veto to be applied. 

\begin{figure}[tb]
\centering
\includegraphics[width=0.49\linewidth]{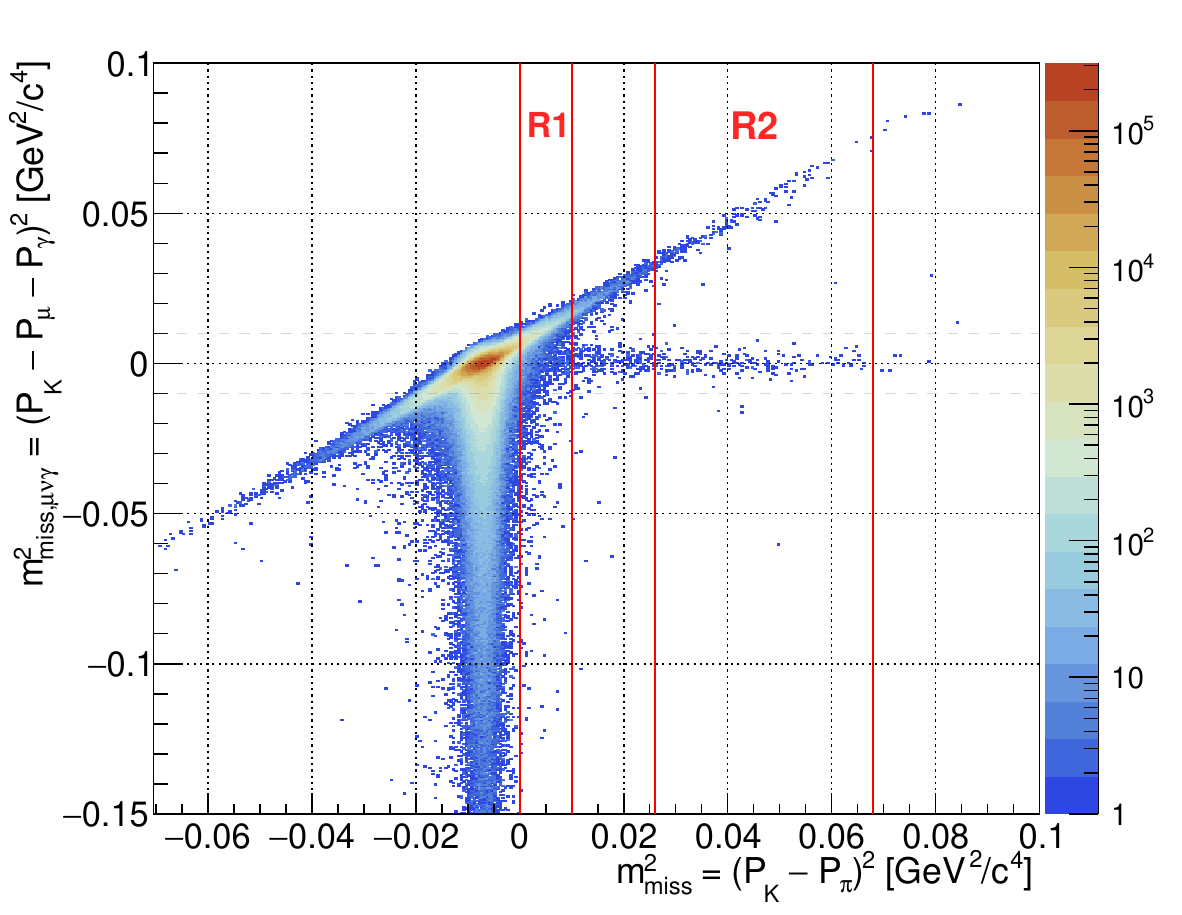}
\includegraphics[width=0.46\linewidth]{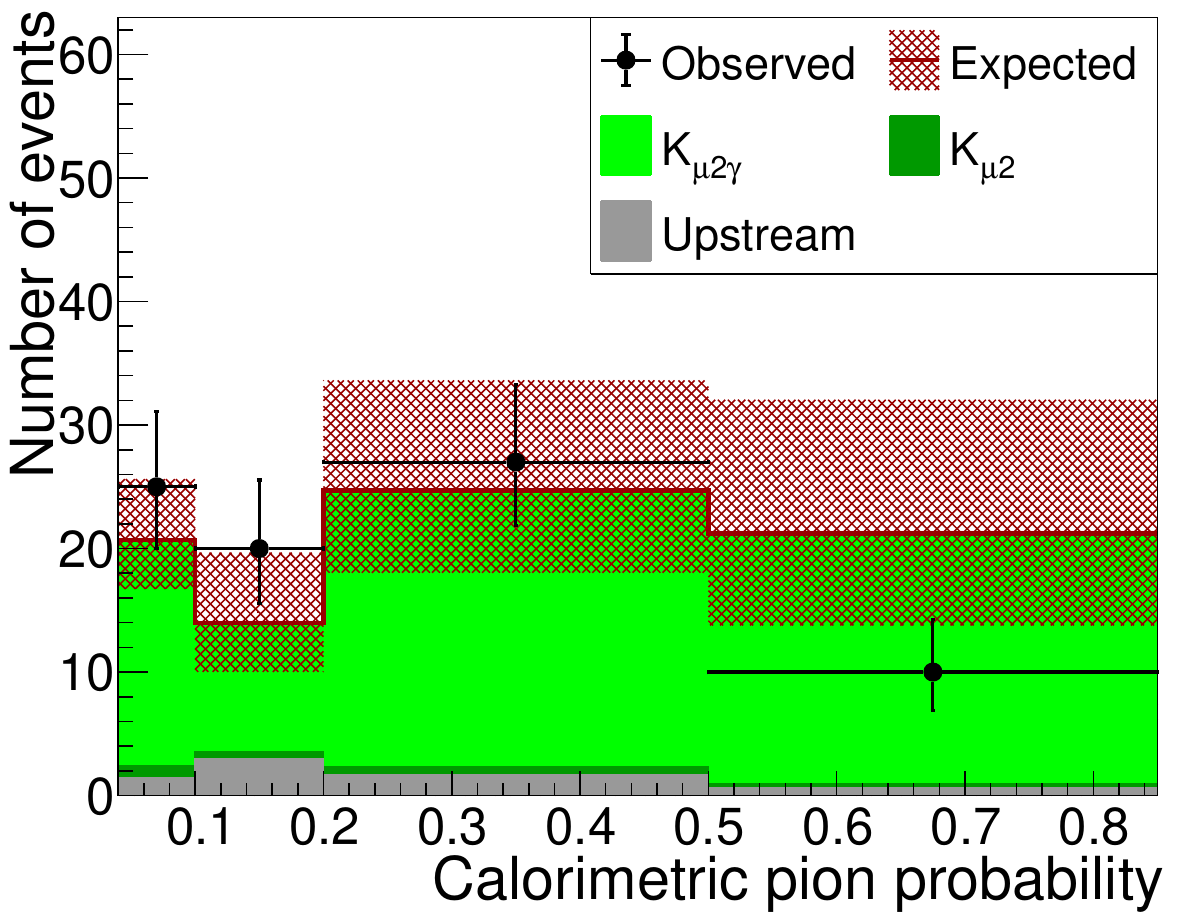}
\caption{
Left: event distribution in the $(m_{\rm miss}^{2},m_{\rm miss,\mu\nu\gamma}^{2})$ plane for a minimum bias data control sample with MUV3 associated signals and no calorimetric BDT condition applied. The $K^{+}\rightarrow\mu^{+}\nu\gamma$ candidates are clearly visible as a horizontal line around $m_{\rm miss,\mu\nu\gamma}^{2}=0$ extending towards high $m_{\rm miss}^{2}$ values including R2.
Right: $K^{+}\rightarrow\mu^{+}\nu\gamma$ background validation samples, four bins in sidebands of the calorimetric BDT pion probability. 
Expectations include contributions from $K^{+}\rightarrow\mu^{+}\nu\gamma$ ($K_{\mu2\gamma}$), $K^{+}\rightarrow\mu^{+}\nu$ ($K_{\mu2}$) and upstream (section~\ref{sec:UpstreamBackground}) events.
}
\label{fig:Kmu2g}
\end{figure}

A squared missing mass variable is used to isolate the $K^{+}\rightarrow\mu^{+}\nu\gamma$ events with $\mu^{+}$ and $\gamma$ overlapping in the LKr:
\begin{equation}
    m_{\rm miss,\mu\nu\gamma}^{2} = (P_{K} - P_{\mu} - P_{\gamma})^{2}\,\,.
\end{equation}
Here $P_{K}$ is the $K^{+}$ 4-momentum measured by the GTK, $P_{\mu}$ is constructed from the STRAW 3-momentum measurement and the $\mu^{+}$ mass, and $P_{\gamma}$ is the 4-momentum of the photon reconstructed using the position and energy $E_{\rm LKr}$ of the LKr cluster (subtracting the nominal $\mu^{+}$ MIP energy deposit, 0.6~GeV), and the decay vertex position.
A $K^{+}\rightarrow\mu^{+}\nu\gamma$ control sample is selected in minimum bias data by applying the $K^{+}\rightarrow\mu^{+}\nu$ kinematic tails selection (including a signal in the MUV3) but without applying any calorimetric BDT constraints: its distribution in the $(m_{\rm miss}^2, m_{\rm miss,\mu\nu\gamma}^{2})$ plane is shown in figure~\ref{fig:Kmu2g}-left. The $K^{+}\rightarrow\mu^{+}\nu\gamma$ events form a peak at $m_{\rm miss,\mu\nu\gamma}^{2} = 0$ with a measured resolution of $2.3\times10^{-3}\,\text{GeV}^{2}/c^{4}$.
Events with $|m_{\rm miss,\mu\nu\gamma}^{2}|<0.01\,\text{GeV}^{2}/c^{4}$, $E_{\rm LKr}>5\,\text{GeV}$, and not satisfying strict $\pi^{+}$ RICH PID criteria are rejected in the signal and normalisation selection. 
This leads to a suppression of the background by a factor of $20$ with a $0.4\%$ relative loss in signal acceptance.

Using the $K^{+}\rightarrow\mu^{+}\nu\gamma$ control sample, the expected background from the process described above is calculated as
\begin{equation}
    N_{b}(K^{+}\rightarrow\mu^{+}\nu\gamma) = N_{\mu\nu\gamma}^\text{MB}\,D_\text{MB}\frac{\varepsilon_\text{PNN}}{\varepsilon_\text{MB}}P_{\rm misID} = 0.82\pm0.43\,\,,
\end{equation}
where: $N_{\mu\nu\gamma}^\text{MB}$ is the number of events in the $K^{+}\rightarrow\mu^{+}\nu\gamma$ control sample with $|m_{\rm miss,\mu\nu\gamma}^{2}|<0.01\,\text{GeV}^{2}/c^{4}$; $D_\text{MB}$ is the downscaling factor of the MB trigger; ${\varepsilon_\text{PNN}}/{\varepsilon_\text{MB}}$ is the ratio of the PNN and MB trigger efficiencies; and $P_{\rm misID}$ is the probability of misidentifying the LKr cluster as one produced by a $\pi^{+}$.
To confirm this prediction a set of validation samples, dominated by $K^{+}\rightarrow\mu^{+}\nu\gamma$ decays, is used. 
These are obtained by applying the signal selection to PNN trigger line data except for the $K^{+}\rightarrow\mu^{+}\nu\gamma$ veto criteria and selecting sidebands in the calorimetric BDT pion probability (figure~\ref{fig:Kmu2g}-right). 

\begin{figure}[tb]
\centering
\includegraphics[width=0.49\linewidth]{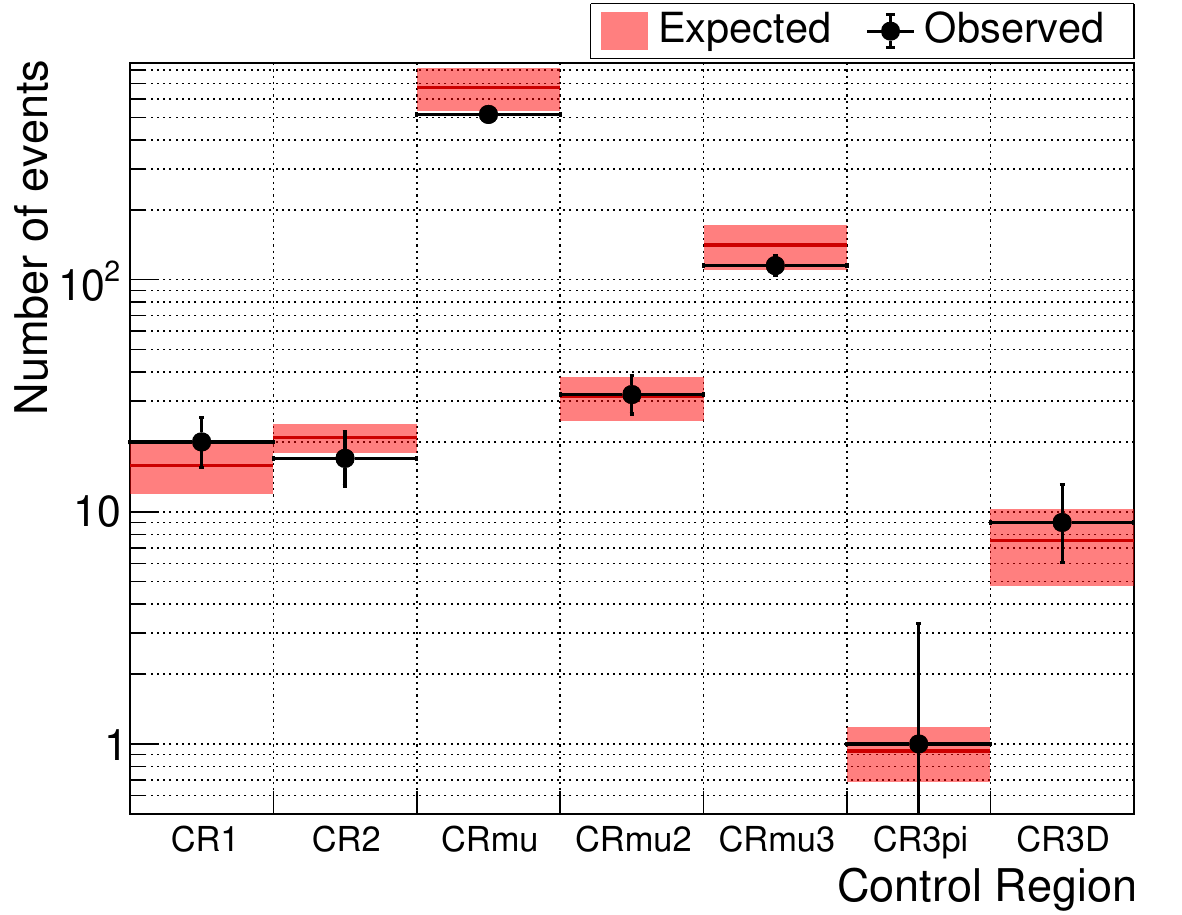}
\includegraphics[width=0.49\linewidth]{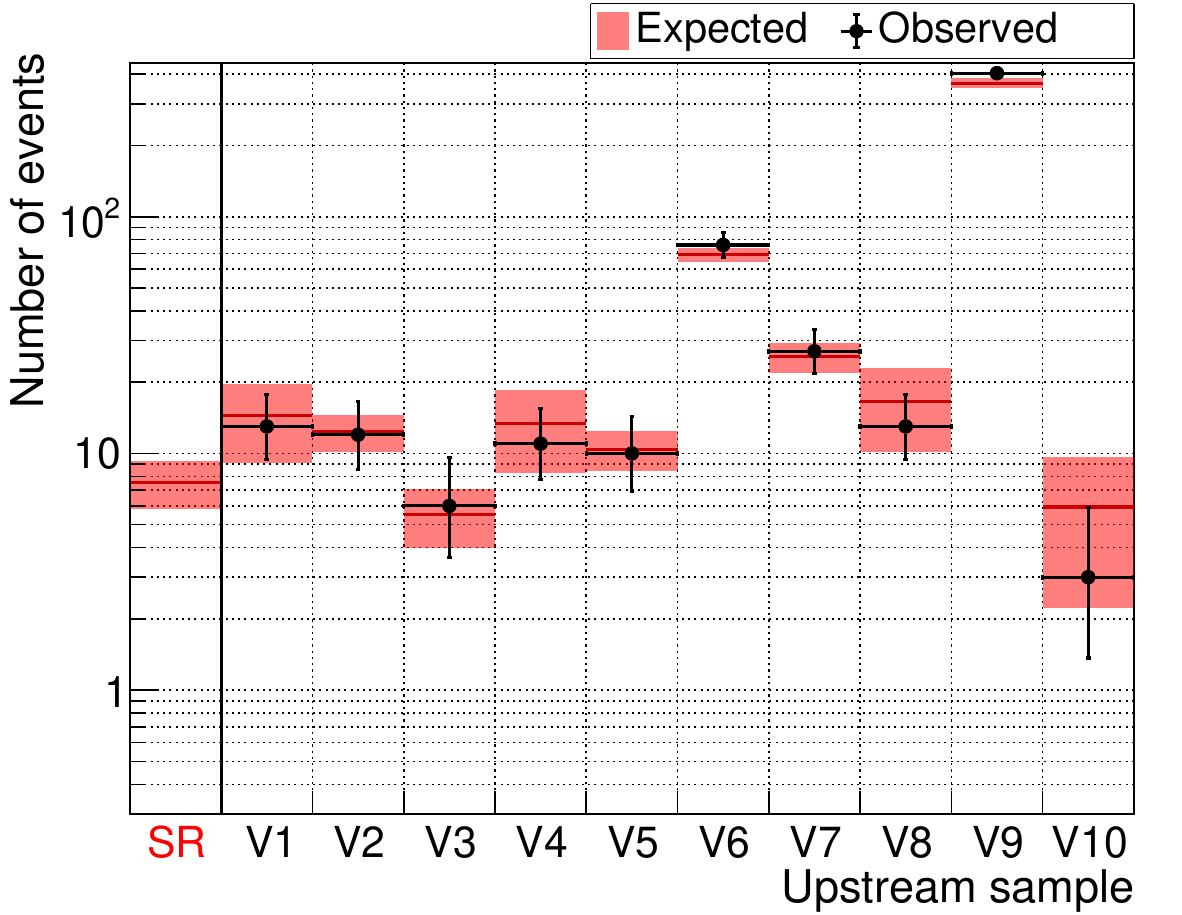}
\caption{
Left: expected and observed numbers of events in the control regions shown in figure~\ref{fig:pi0Rejection}-right. 
The global $p$-value of the comparison is 0.80, the lowest single-region $p$-value is 0.24.
Right: expected and observed numbers of events in the upstream validation samples.  
The first bin 
is the signal region and therefore only the expectation is displayed. The global $p$-value is 0.97, the lowest single-sample $p$-value is 0.14.
}
\label{fig:ControlAndValidRegions}
\end{figure}

Background estimates for the main kaon decay modes in the FV are validated using the control regions defined in the $(p_{\pi^{+}},m_{\rm miss}^{2})$ plane, such that they are primarily populated by the relevant kaon decays (figure~\ref{fig:pi0Rejection}-right). Results are shown in figure~\ref{fig:ControlAndValidRegions}-left.

\subsection{Other $K^+$ decay backgrounds}
\label{sec:BkgFromSimulations}
For other backgrounds from $K^{+}$ decays in the FV there are no clean control samples in data, therefore simulations are used to evaluate background expectations. 
The primary background of this type is from $K^{+}\rightarrow\pi^{+}\pi^{-}e^{+}\nu$ decays.
From a sample of $2\times10^{9}$ simulated events, the overall acceptance is 
$A_{\pi\pi e\nu} = (1.3\pm0.3)\times10^{-8}$
and the background expectation is
\begin{equation}
    N_{b}(K^{+}\rightarrow\pi^{+}\pi^{-}e^{+}\nu) = N_{K}\,\varepsilon_\text{RV}\,\varepsilon_\text{trig}\,\mathcal{B}({K^{+}\rightarrow\pi^{+}\pi^{-}e^{+}\nu})\,A_{\pi\pi e\nu}
    = 0.89^{+0.33}_{-0.27}\,. 
\end{equation}
The asymmetry in the uncertainty arises from the small numbers of simulated $K^{+}\rightarrow\pi^{+}\pi^{-}e^{+}\nu$ events that satisfy the signal selection in individual $\pi^{+}$ momentum bins.

Other $K^{+}$ decay backgrounds are found to be negligible. 
The largest of these is from $K^{+}\rightarrow\pi^{+}\gamma\gamma$ decays, estimated to be $0.01\pm0.01$. 
The next largest is from $K^{+}\rightarrow\pi^{0}\ell^{+}\nu$ decays, estimated to be less than $10^{-3}$.

\subsection{Upstream background}
\label{sec:UpstreamBackground}
Upstream background events populate the signal region if a decay or beam interaction upstream of the FV produces a $\pi^{+}$ which is detected downstream, and a fake vertex is reconstructed in the FV.
This constitutes the largest background, and is estimated with a data-driven strategy.

\begin{figure}[tb]
\centering
\includegraphics[width=0.49\linewidth]{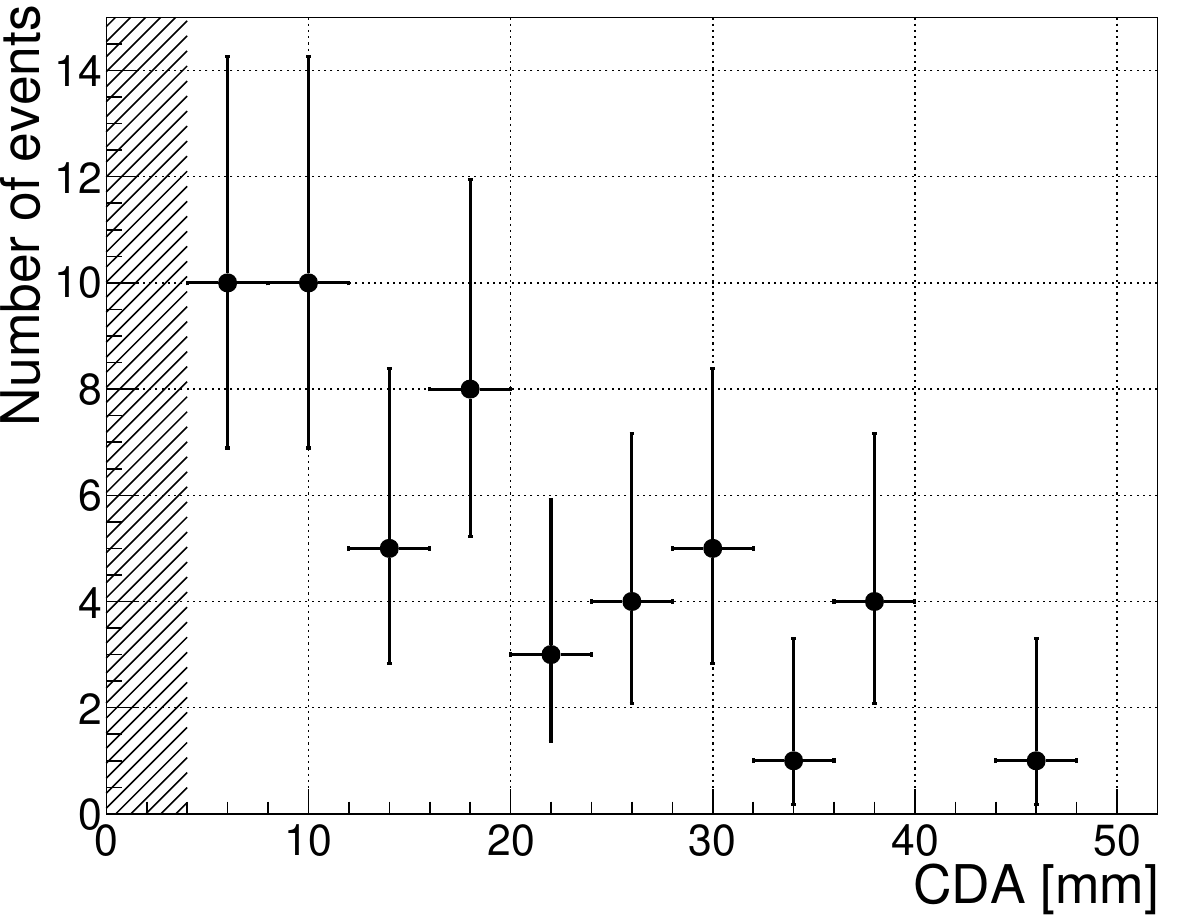}
\includegraphics[width=0.49\linewidth]{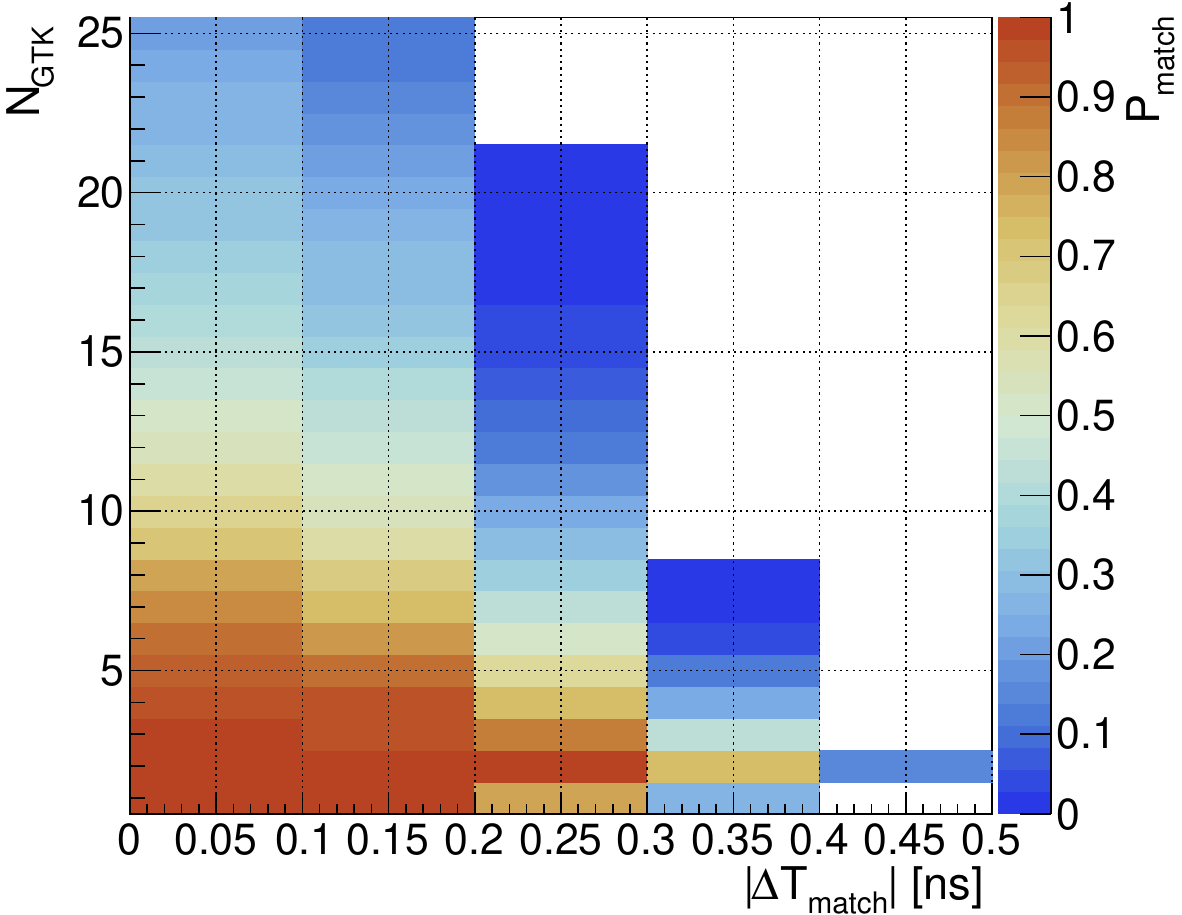}
\caption{
Left: distribution of the CDA variable for events in the upstream reference sample (URS).
Right: matching probability $P_{\text{match}}$ measured in the $K^{+}\rightarrow\pi^{+}\pi^{0}$ normalisation sample in bins of $\Delta T_{\text{match}}$ and $N_{\text{GTK}}$.
}
\label{fig:Upstream}
\end{figure}

An upstream reference sample (URS) is selected by applying the full signal selection except for the $K^{+}$--$\pi^{+}$ matching criteria and requiring $\text{CDA}>4\,\text{mm}$. The URS contains $N_{\rm URS} = 51$ events from all types of upstream background. 
A factor $f_{\rm CDA}$ is used to extrapolate the CDA distribution of the URS (figure~\ref{fig:Upstream}-left) to the signal region ($\text{CDA}<4$~mm).
Assuming that the distribution is flat for $\text{CDA}<12$~mm (validated within uncertainties in alternative samples) leads to $f_{\rm CDA} = 0.20\pm0.03$.
The probability of an event satisfying the $K^{+}$--$\pi^{+}$ matching criteria based on the Bayesian discriminant described in section~\ref{sec:NormSelection}, $P_{\text{match}}(\Delta T_{\text{match}},N_{\text{GTK}})$, is measured using $K^{+}\rightarrow\pi^{+}\pi^{0}$ normalisation data events, weighted to produce a flat CDA distribution between 0 and 4~mm.  
Results are given in figure~\ref{fig:Upstream}-right.
Finally, the upstream background expectation is evaluated as
\begin{equation}
    N_{b}(\text{Upstream}) = 
    f_{\rm CDA} \sum_{i} N^{\rm URS}_{i} P_{i}^{\text{match}}  
    = 7.4^{+2.1}_{-1.8}\,\,,
\end{equation}
where the sum runs over the 2-dimensional $(\Delta T_{\text{match}},N_{\text{GTK}})$ bins shown in figure~\ref{fig:Upstream}-right.
The average matching probability, given the distribution of URS events, is 73\%.
The uncertainty quoted is primarily statistical, 
with a sub-leading systematic contribution due to the assumption of flatness of the CDA distribution.

The upstream background estimate is validated using a set of samples, defined by loosening and inverting individual upstream veto conditions 
to enhance certain mechanisms responsible for particular upstream backgrounds.
Interaction-enriched samples V1 and V7 are defined by selecting events with $\pi^{+}$ pointing to the candidate $K^{+}$ position at the GTK3 and events with signals in the CHANTI, respectively. 
Accidental-enriched samples V3, V5 and V9 are defined by inverting GTK pileup rejection, VC conditions and upstream BDT criteria, respectively. 
The samples V1, V3, V5, V7 and V9 require events to be in the kinematic signal regions; samples V2, V4, V6, V8 and V10 are defined similarly except selecting the kinematic region $m_{\rm miss}^{2}<-0.05\,\text{GeV}^{2}/c^{4}$, which does not include well-reconstructed $K^{+}$ decays in the FV.
Results of the validation are shown in figure~\ref{fig:ControlAndValidRegions}-right.
All these samples are statistically independent, and the good agreement across them validates the background evaluation procedure.
The VC is essential to control the upstream background, reducing the background expectation by a factor of 2. 
Removing the VC conditions from the signal selection leads to $9$ additional observed events, in agreement with the prediction of $6.9\pm1.4$.

A summary of the background expectations is given in table~\ref{tab:BackgroundsSummaryTable}.

\begin{table} 
    \centering
    \caption{Background expectations for 2021--2022 data, summed over the six $\pi^{+}$ momentum bins.}
    \vspace{8pt}
    \begin{tabular}{|l|l|} 
    \hline
    Background & Events \T\B \\
    \hline
    \hline 
    $K^{+}\rightarrow\pi^{+}\pi^{0}(\gamma)$ & ~$0.83\pm0.05$  \T\B \\ 
    $K^{+}\rightarrow\mu^{+}\nu(\gamma)$ & ~$1.70\pm0.47$  \T\B \\ 
    $K^{+}\rightarrow\pi^{+}\pi^{+}\pi^{-}$ & ~$0.11\pm0.03$ \T\B \\
    $K^{+}\rightarrow\pi^{+}\pi^{-}e^{+}\nu$ & ~$0.89^{+0.33}_{-0.27}$ \T\B \\
    $K^{+}\rightarrow\pi^{+}\gamma\gamma$ & ~$0.01\pm0.01$ \T\B \\
    $K^{+}\rightarrow\pi^{0}\ell^{+}\nu$ & ~$<0.001$  \T\B \\ 
    Upstream & ~$7.4^{+2.1}_{-1.8}$  \T\B \\
    \hline
    Total & \hspace{-2pt}$11.0^{+2.1}_{-1.9}$ \T\B \\
    \hline
    \end{tabular} 
    \label{tab:BackgroundsSummaryTable}
\end{table}

\section{Results}
\label{sec:Results}

\subsection{Data sample 2021--2022}
\label{sec:Results2122}
\begin{figure}[tb]
\centering
\includegraphics[width=0.49\linewidth]{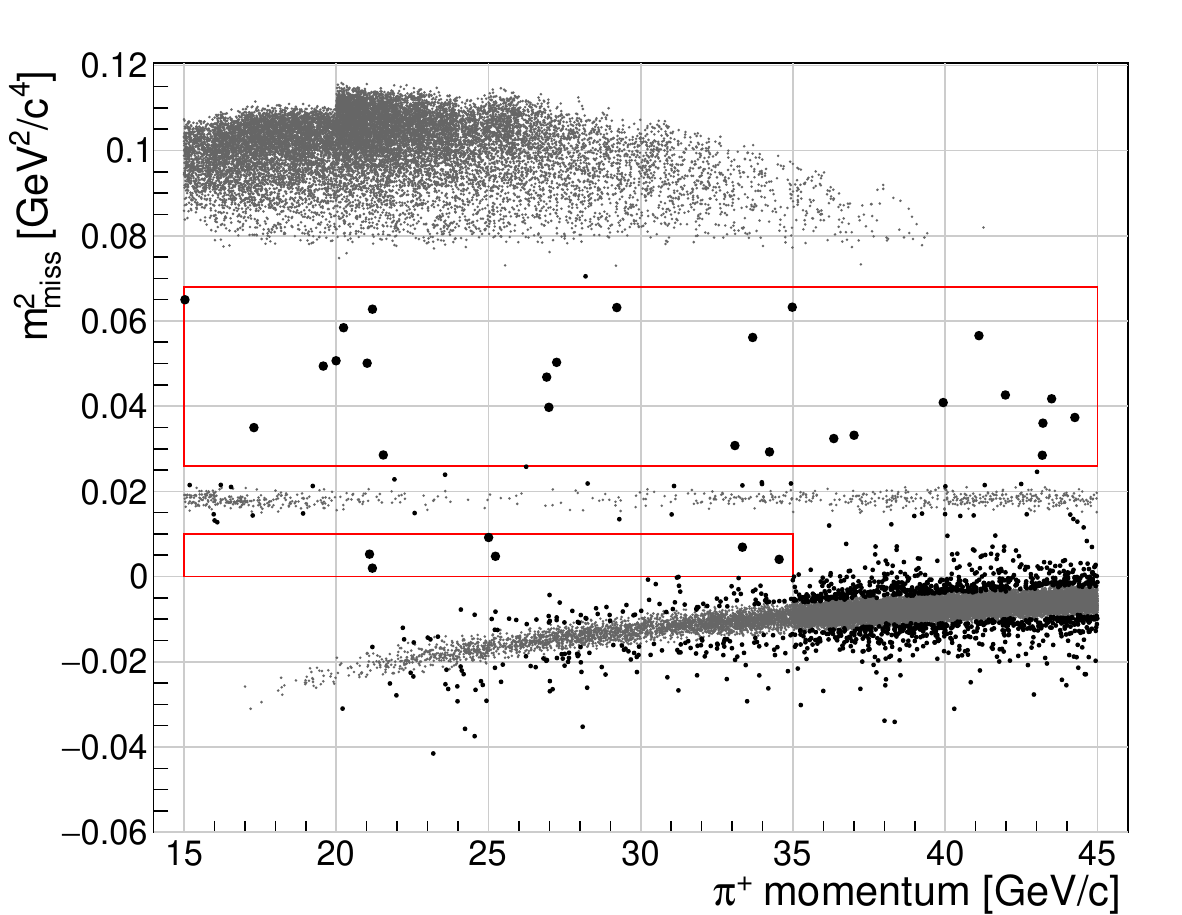}
\includegraphics[width=0.47\linewidth]{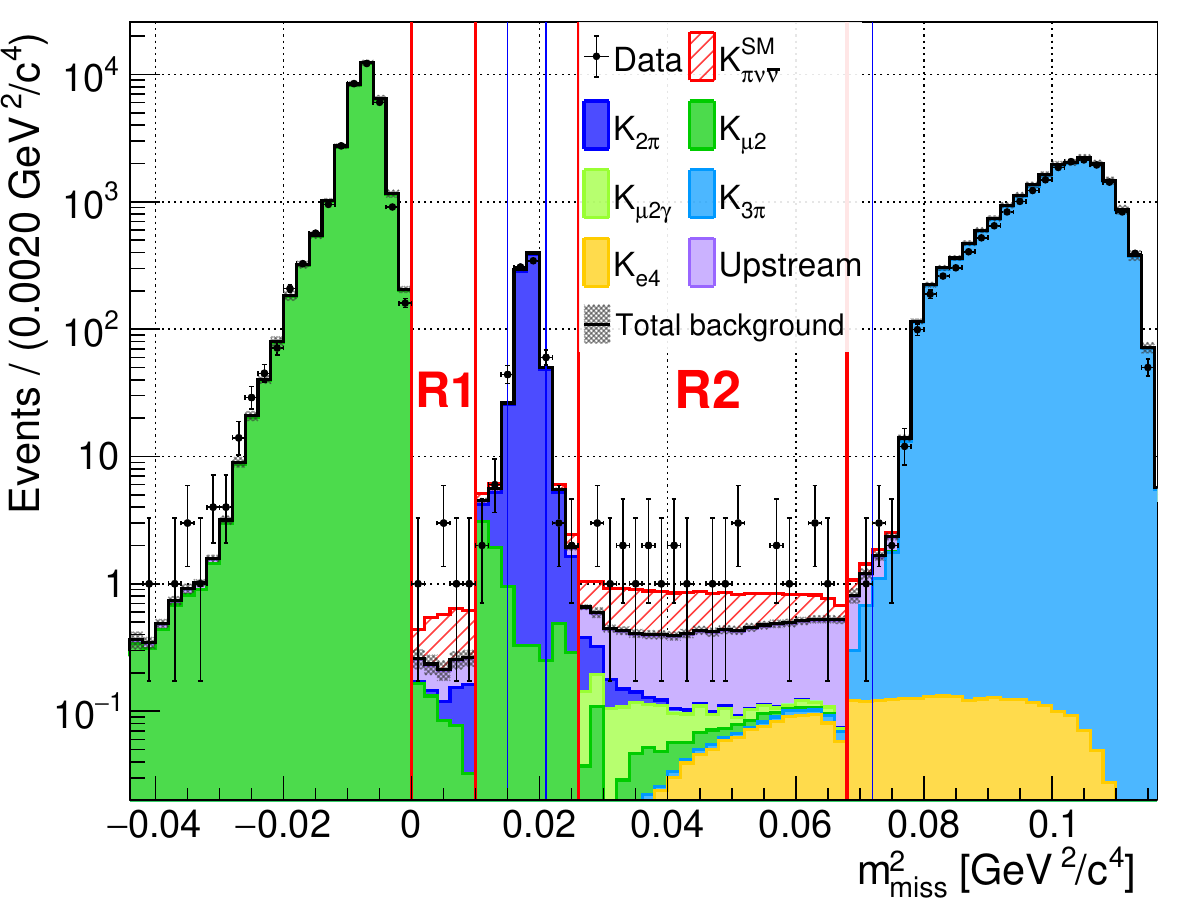}
\caption{
Left: distribution of the observed data events satisfying the signal selection criteria in the $(p_{\pi^{+}},m_{\rm miss}^{2})$ plane. 
Events in the background, control and signal regions are shown by small grey, small black and large black markers, respectively.
Right:  $m_{\rm miss}^{2}$ projection including 
SM signal~\cite{Buras:2015qea} (assuming $\mathcal{B}_{\pi\nu\bar{\nu}}^{\text{SM}} = 8.4\times10^{-11}$) and background expectations from $K^{+}\to\pi^{+}\pi^{0}$ ($K_{2\pi}$), $K^{+}\to\mu^+\nu$ ($K_{\mu2}$), $K^{+}\to\mu^+\nu\gamma$ ($K_{\mu2\gamma}$), $K^{+}\to\pi^{+}\pi^{+}\pi^{-}$ ($K_{3\pi}$), $K^+\to\pi^+\pi^-e^+\nu$ ($K_{e4}$)  
and upstream.
The total expected background and its uncertainty is shown by the black line and hatched bars, respectively.
In the signal region R1, events in the momentum range $35$--$45\,\text{GeV}/c$ are excluded.
}
\label{fig:Observed}
\end{figure}

Figure~\ref{fig:Observed}-left shows the distribution of the observed data events satisfying the signal selection criteria in the $(p_{\pi^{+}},m_{\rm miss}^{2})$ plane. 
In total, 6 events are observed in R1 and 25 in R2. 
The $m_{\rm miss}^{2}$ projection including the background spectra and SM signal expectation is shown in figure~\ref{fig:Observed}-right.

The branching ratio measurement is performed using a profile likelihood ratio test statistic 
\begin{equation}
    q(\theta|\mathbf{n};\boldsymbol{\nu}) = -2\,\ln\left( \frac{L(\theta;\hat{\hat{\boldsymbol\nu}})}{L(\hat{\theta};\hat{\boldsymbol\nu})} \right) \,\,,
\end{equation}
where $\theta = \mathcal{B}_{\pi\nu\bar\nu} / \mathcal{B}_{\pi\nu\bar{\nu}}^{\text{SM}}$ (signal strength) is the parameter of interest and the nuisance parameters $\boldsymbol{\nu}$ take into account the uncertainties in the signal
and background expectations.
The analysis is performed using $N_{\text{cat}}$ categories, considered independent. The likelihood function $L(\theta;\boldsymbol\nu)$ takes into account the Poissonian fluctuations of the observed counts, the Gaussian uncertainty of the signal expectation, and the asymmetric uncertainty of the background estimate. 
The best fit value of $\theta$ is at the minimum of the function $q(\theta)$, with the one standard deviation range ($68\%$ confidence interval) defined by $q(\theta)<1$. 
The statistical uncertainty is evaluated by performing a similar procedure but assuming that the signal and background expectations are known exactly, and therefore using as a test statistic
$q^{\prime}(\theta)=-2\,\ln{(L^\prime(\theta)/L^\prime(\hat{\theta}))}$, where $L^\prime$ is the likelihood function in the hypothesis of independent Poisson distributed observations~\cite{PDG}.
The systematic uncertainty is derived as the contribution to be added in quadrature with the statistical uncertainty to reach the total uncertainty.

\begin{figure}[tb]
\centering
\includegraphics[width=0.49\linewidth]{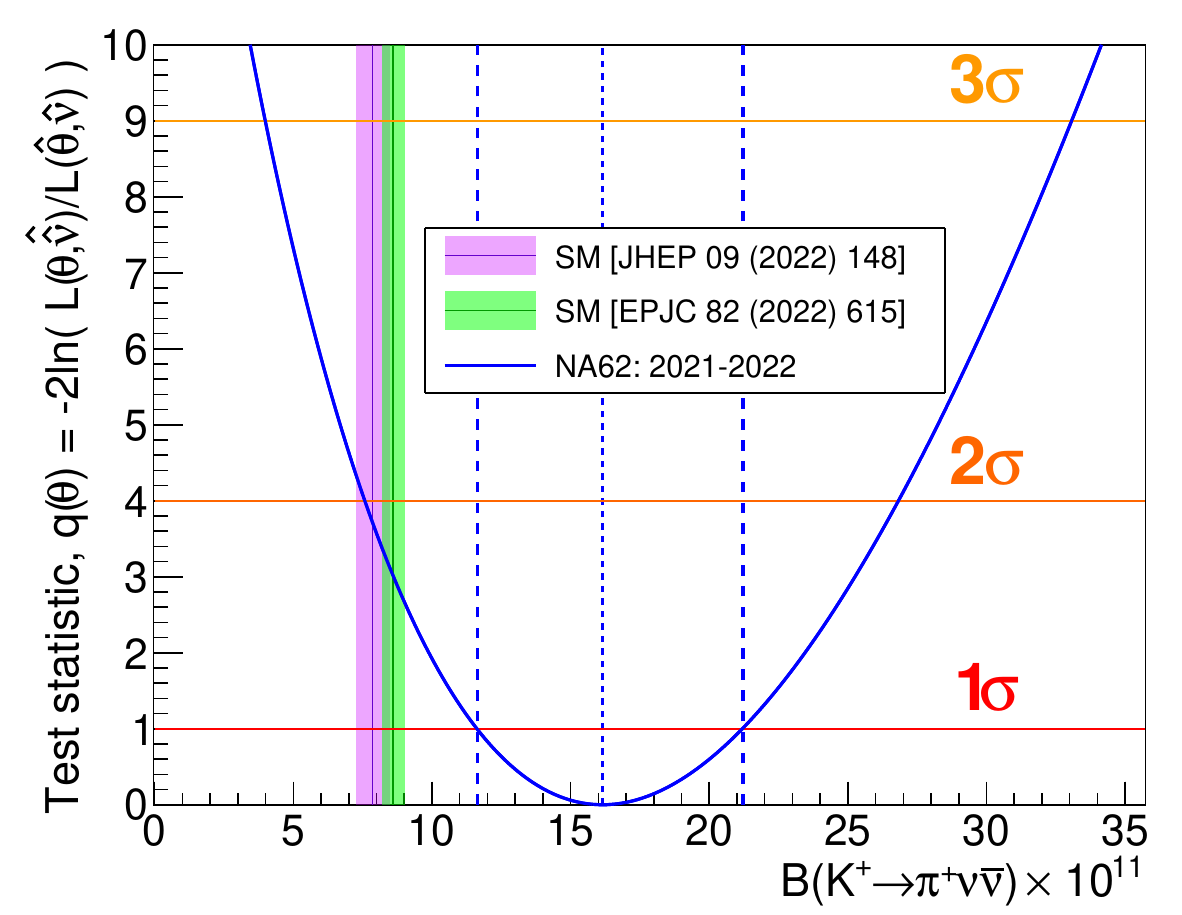}
\includegraphics[width=0.49\linewidth]{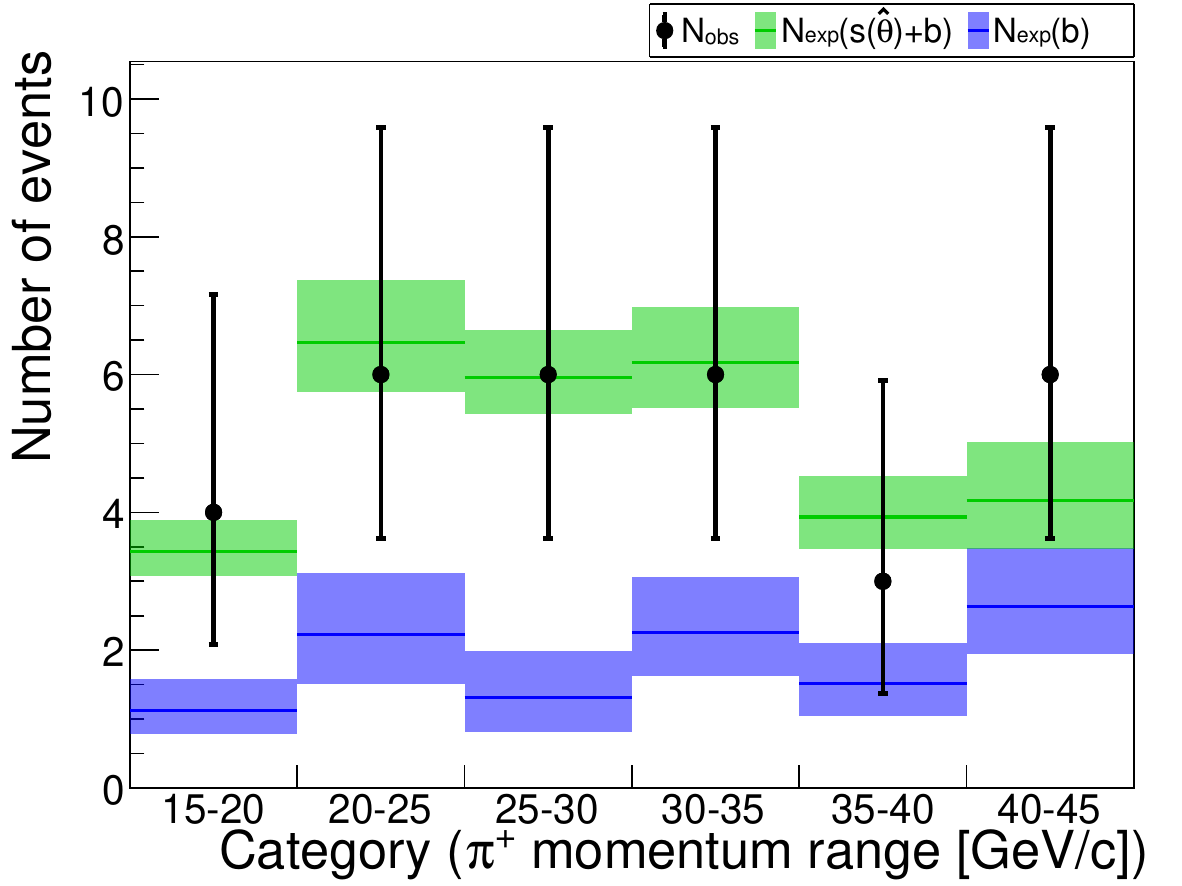}
\caption{
Left: test statistic $q$ as a function of the $K^{+}\rightarrow\pi^{+}\nu\bar{\nu}$ branching ratio for 2021--2022 data. 
Right: numbers of expected and observed events in the six categories used for the statistical analysis of 2021--2022 data. The background expectation is shown in blue, while the signal (using the measured value of the branching ratio) plus background expectation is shown in green.
}
\label{fig:StatResults2122}
\end{figure}

The analysis of 2021--2022 data is performed using $N_{\text{cat}}=6$, corresponding to the six $\pi^+$ momentum bins defined above.
The test statistic is displayed as a function of the branching ratio in figure~\ref{fig:StatResults2122}-left. The resulting measurement of the branching ratio is
\begin{equation}
\begin{aligned}
    \mathcal{B}_{2021-2022}(K^{+}\rightarrow\pi^{+}\nu\bar{\nu}) 
&= 
\left(
16.2
\left. {}^{+4.9}_{-4.3} \right|_{\text{stat}}
\left. {}^{+1.4}_{-1.4} \right|_{\text{syst}}
\right)
\times10^{-11}
\\&=
\left(
16.2^{+5.1}_{-4.5}\right)\times10^{-11} 
\,\,.
\end{aligned}
\label{eqn:BR2122}
\end{equation}
The comparison between the numbers of events observed and expected in each category is shown in figure~\ref{fig:StatResults2122}-right, where the expectations are based on the measured value of the branching ratio.
The goodness of fit is quantified by $q^{\prime}_{\text{min}} / \text{ndf} = 1.1/5$, where $q^{\prime}_{\text{min}}$ is the minimum value of $q^{\prime}(\theta)$, 
and $\text{ndf} = (N_{\text{cat}}-1)$ is the number of degrees of freedom.

\subsection{Combination of data samples 2016--2022}
\label{sec:Results1622}
The six categories of the 2021--2022 data have been combined with the nine categories spanning the 2016--2018 data~\cite{Pnn2016paper, Pnn2017paper,PnnRun1Paper} for a more precise measurement of the branching ratio.
The numbers of observed and expected SM signal and background events for each category are summarised in table~\ref{tab:StatTreatmentInputTable}.

\begin{table}
\caption{
Inputs to the statistical combination of the 2016--2022 data: analysis category, data sample, $\pi^+$ momentum range, numbers of observed and expected SM signal and background events.
Results from the 2016~\cite{Pnn2016paper}, 2017~\cite{Pnn2017paper} and 2018~\cite{PnnRun1Paper} data  are combined with those from the present 2021--2022 data analysis. Using equation~\ref{eqn:Npnn}, $N_{\pi\nu\bar{\nu}}^{\text{SM}}$ is evaluated assuming $\mathcal{B}_{\pi\nu\bar{\nu}}^{\text{SM}} = 8.4\times10^{-11}$.
}
\vspace{8pt}
\centering
\begin{tabular}{|c|c|c||c||c|c|}
\hline 
Category & Sample & $p_{\pi^{+}}$ range ($\text{GeV}/c$) & $N_\text{obs}$ & $N_{\pi\nu\bar{\nu}}^{\text{SM}}$ & $N_{b}$\tabularnewline
\hline 
\hline 
1 & 2016 & $15$--$35$ & $1$ & $0.267\pm0.020$ & $0.152_{-0.035}^{+0.093}$\tabularnewline
\hline 
2 & 2017 & $15$--$35$ & $2$ & $2.16\pm0.13$ & $1.46\pm0.33$\tabularnewline
\hline 
3 & 2018 S1 & $15$--$45$ & $2$ & $1.56\pm0.10$ & $1.11_{-0.22}^{+0.40}$\tabularnewline
\hline 
4 & \multirow{6}{*}{2018 S2} & $15$--$20$ & $1$ & $0.56\pm0.04$ & $1.14_{-0.30}^{+0.78}$\tabularnewline
5 &  & $20$--$25$ & $4$ & $1.43\pm0.09$ & $1.02_{-0.28}^{+0.67}$\tabularnewline
6 &  & $25$--$30$ & $2$ & $1.53\pm0.10$ & $0.41_{-0.10}^{+0.32}$\tabularnewline
7 &  & $30$--$35$ & $6$ & $1.32\pm0.09$ & $1.09_{-0.30}^{+0.52}$\tabularnewline
8 &  & $35$--$40$ & $1$ & $0.69\pm0.04$ & $0.29_{-0.10}^{+0.31}$\tabularnewline
9 &  & $40$--$45$ & $1$ & $0.48\pm0.03$ & $0.35_{-0.12}^{+0.41}$\tabularnewline
\hline 
10 & \multirow{6}{*}{2021--2022} & $15$--$20$ & $4$ & $1.20\pm0.04$ & $1.12_{-0.34}^{+0.46}$\tabularnewline
11 &  & $20$--$25$ & $6$ & $2.21\pm0.07$ & $2.23_{-0.71}^{+0.90}$\tabularnewline
12 &  & $25$--$30$ & $6$ & $2.41\pm0.07$ & $1.32_{-0.51}^{+0.68}$\tabularnewline
13 &  & $30$--$35$ & $6$ & $2.03\pm0.06$ & $2.26_{-0.64}^{+0.81}$\tabularnewline
14 &  & $35$--$40$ & $3$ & $1.26\pm0.04$ & $1.51_{-0.46}^{+0.59}$\tabularnewline
15 &  & $40$--$45$ & $6$ & $0.80\pm0.03$ & $2.63_{-0.69}^{+0.86}$\tabularnewline
\hline

\end{tabular}
\label{tab:StatTreatmentInputTable}
\end{table}

\begin{figure}[tb]
\centering
\includegraphics[width=0.49\linewidth]{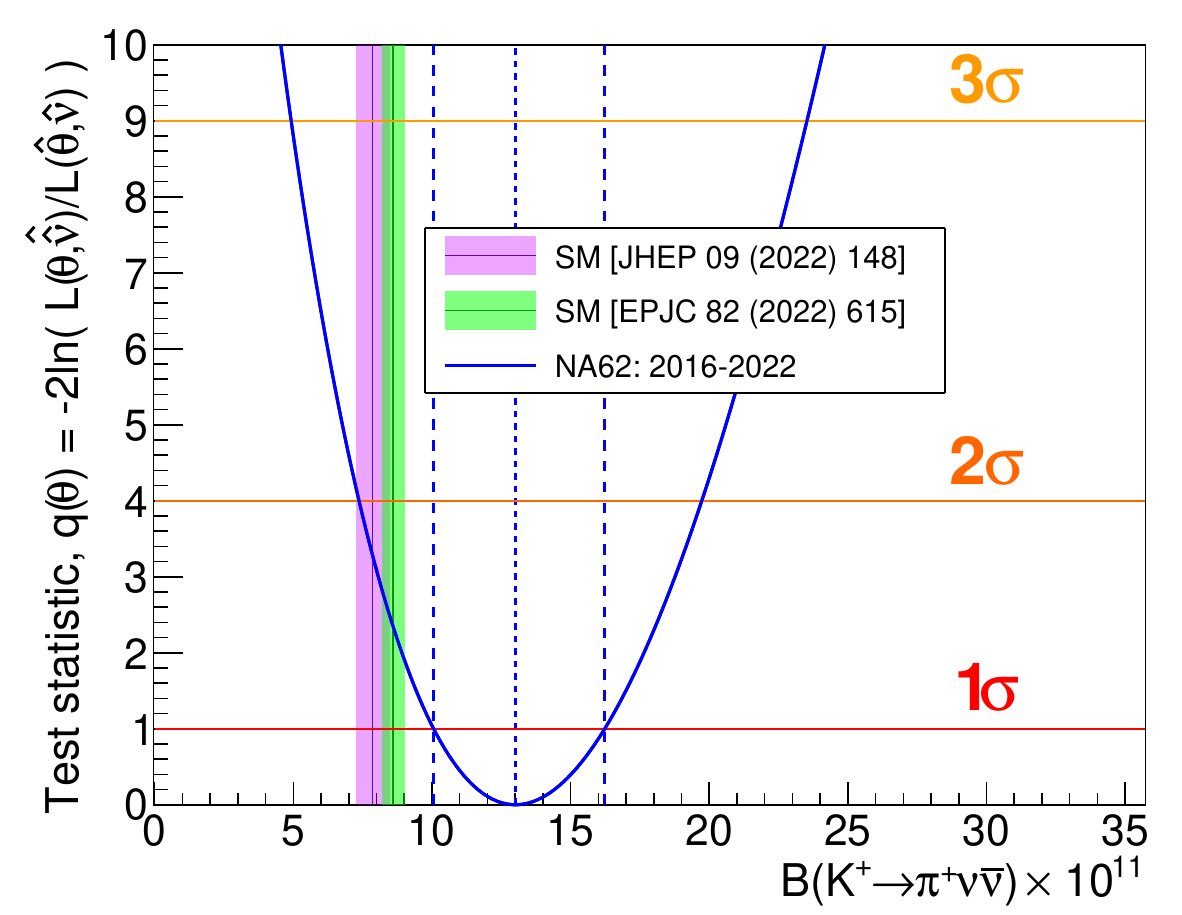}
\includegraphics[width=0.49\linewidth]{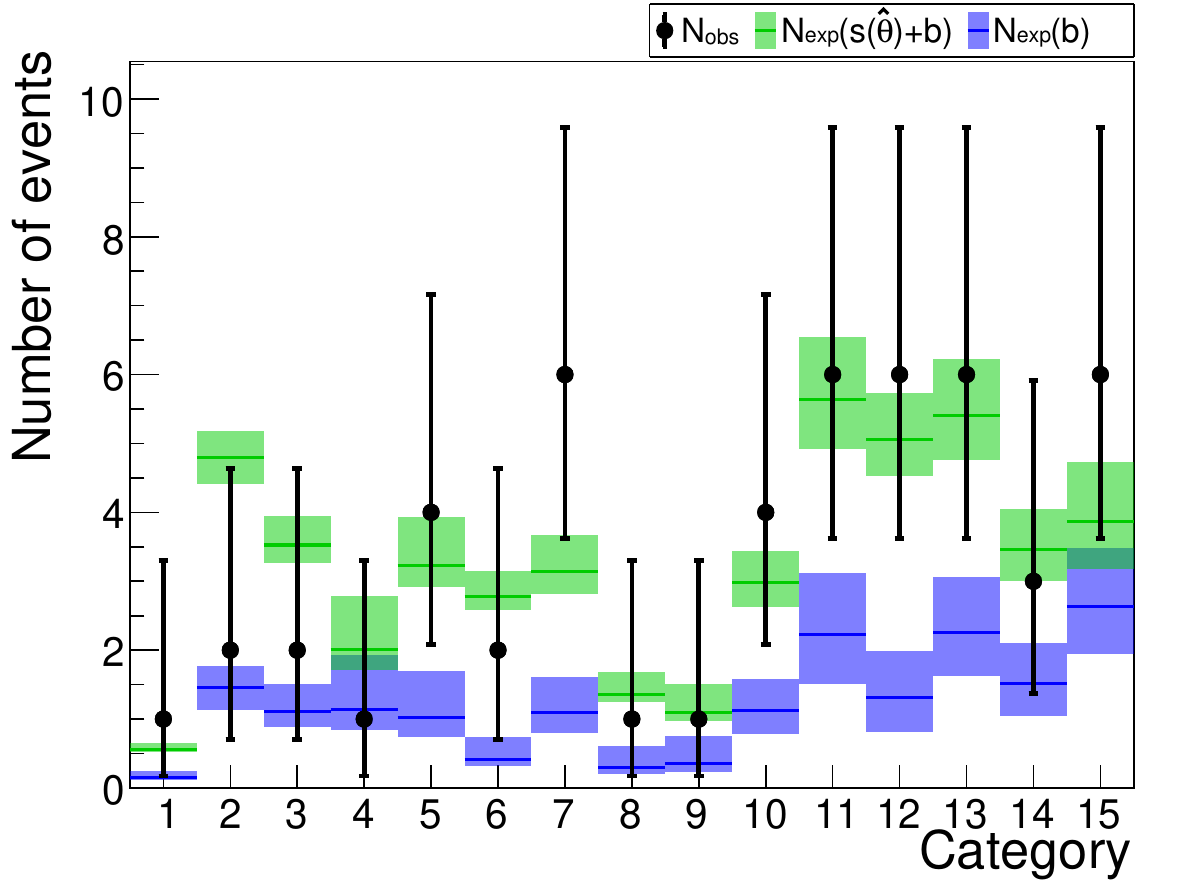}
\caption{
Left: test statistic $q$ as a function of the $K^{+}\rightarrow\pi^{+}\nu\bar{\nu}$ branching ratio for 2016--2022 data.
Right: numbers of expected and observed events in the 15 categories used for the statistical analysis of 2016--2022 data
(table~\ref{tab:StatTreatmentInputTable}).
The background expectation is shown in blue, while the signal (using the measured value of the branching ratio) plus background expectation is shown in green.
}
\label{fig:StatResults1622}
\end{figure}

The combined branching ratio measurement is obtained using the procedure described in section~\ref{sec:Results2122}. 
The resulting test statistic as a function of the branching ratio is shown in figure~\ref{fig:StatResults1622}-left. 
The comparison of expectation (based on the measured branching ratio value) and observations across the 15 categories is shown in figure~\ref{fig:StatResults1622}-right, with $q^{\prime}_{\text{min}} / \text{ndf} = 8.0/14$. 
The result is
\begin{equation}
\begin{aligned}
    \mathcal{B}_{2016-2022}(K^{+}\rightarrow\pi^{+}\nu\bar{\nu}) 
&=
\left(
13.0
\left.{}^{+ 3.0}_{ - 2.7} \right|_{\text{stat}} \left.{}^{+ 1.3}_{- 1.3} \right|_{\text{syst}}  \right)\times10^{-11}
\\
&= 
\left(13.0^{+ 3.3}_{- 3.0} \right)\times10^{-11} 
\,\,.
\end{aligned}
\label{eqn:BR1622}
\end{equation}
This result includes an additional systematic uncertainty of $0.5\times10^{-11}$, which arises from the combination of the
different datasets, analysed with different strategies, in particular with different background estimation procedures.

For the full 2016--2022 dataset, with an expectation of $18^{+3}_{-2}$ background events and an observation of $51$ events, the $p$-value of  the background-only hypothesis is evaluated to be $2\times10^{-7}$.
Therefore, for the first time, the background-only hypothesis is rejected with a significance above $5\,\sigma$, which marks the first observation of the $K^{+}\rightarrow\pi^{+}\nu\bar{\nu}$ decay. 

\begin{figure}[tb]
\centering
\includegraphics[width=0.49\linewidth]{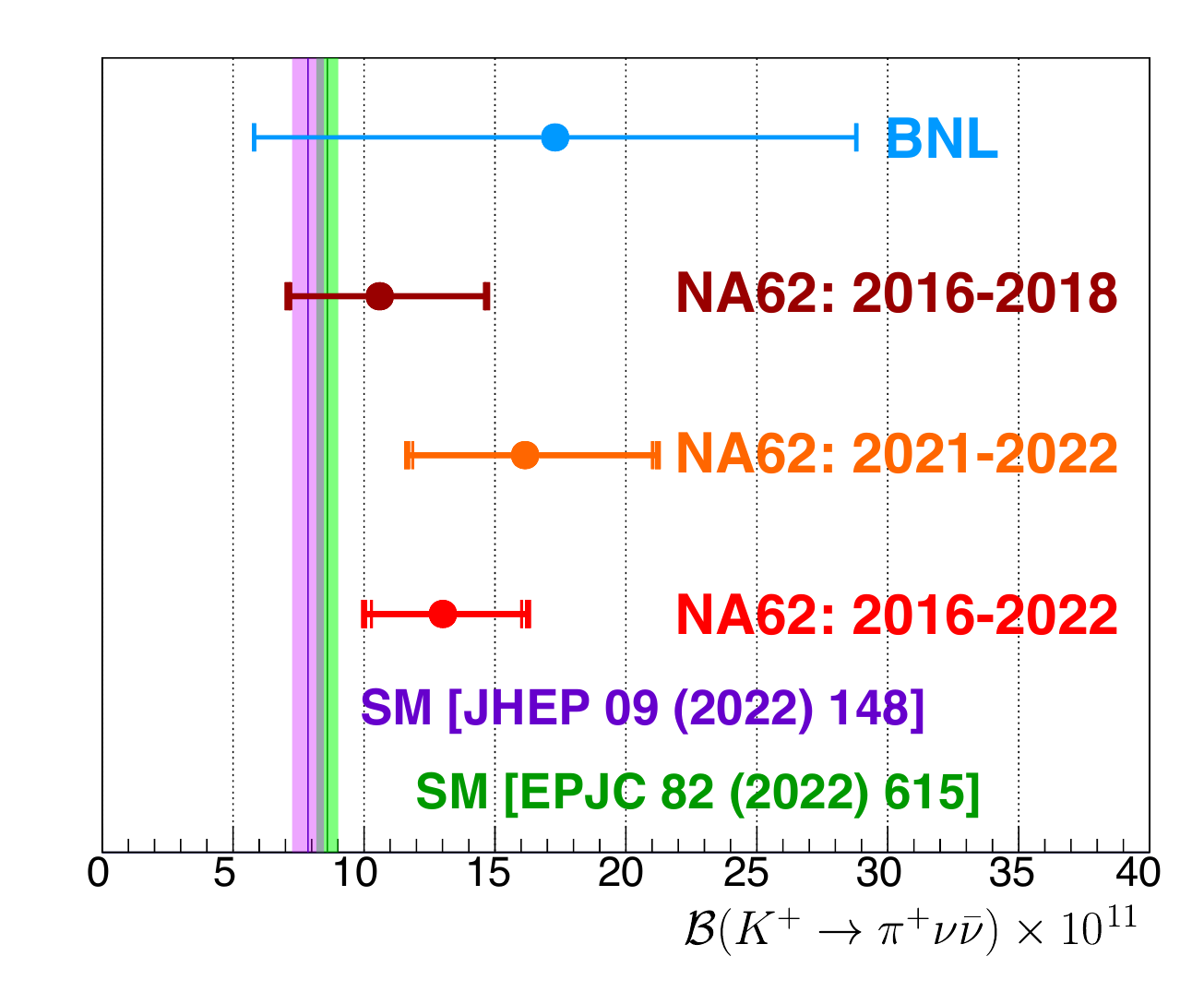}
\includegraphics[width=0.49\linewidth]{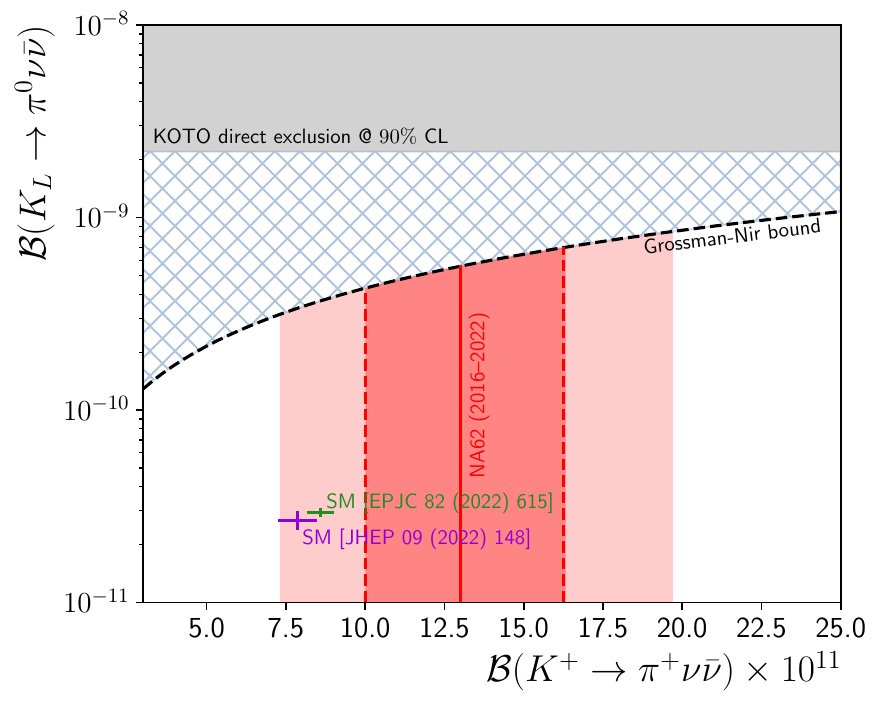}
\caption{
Left: summary of $K^{+}\rightarrow\pi^{+}\nu\bar{\nu}$ branching ratio measurements 
from the BNL E787 and E949 experiments~\cite{BNL-E949:2009dza}, 
and the NA62 experiment using the 2016--2018~\cite{PnnRun1Paper}, 2021--2022 (equation~\ref{eqn:BR2122}) and 2016--2022 (equation~\ref{eqn:BR1622}) data.
Statistical and total uncertainties are shown by thinner and thicker vertical bars, respectively.
These are compared to the two recent SM predictions~\cite{Buras:2022wpw,DAmbrosio:2022kvb}.
Right: global status of the $K\rightarrow\pi\nu\bar{\nu}$ decay modes, showing the most stringent $\mathcal{B}(K_{L}\rightarrow\pi^{0}\nu\bar{\nu})$ upper limit~\cite{KOTO:2024zbl}, 
the Grossman-Nir bound~\cite{Grossman:1997sk,PDG}, the two recent SM predictions~\cite{Buras:2022wpw,DAmbrosio:2022kvb}, and the $\mathcal{B}(K^{+}\rightarrow\pi^{+}\nu\bar{\nu})$ result from the combined 2016--2022 NA62 dataset (the $1\,\sigma$ and $2\,\sigma$ ranges are displayed in darker and lighter shaded areas, respectively).
}
\label{fig:ResultsInContext}
\end{figure}

The $K^{+}\rightarrow\pi^{+}\nu\bar{
\nu}$ branching ratio measurements and the updated experimental and theoretical status are summarised in figure~\ref{fig:ResultsInContext}.

\section{Conclusions}

The $K^{+}\rightarrow\pi^{+}\nu\bar{\nu}$ decay is observed with a significance above $5\,\sigma$, 
and its branching ratio is measured to be $\mathcal{B}(K^{+}\rightarrow\pi^{+}\nu\bar{\nu}) = \left( 13.0^{+ 3.3}_{- 3.0} \right)\times10^{-11}$.
As a result, $\mathcal{B}(K^{+}\rightarrow\pi^{+}\nu\bar{\nu})$ becomes the smallest branching ratio measured with a signal significance above $5\,\sigma$.
The relative precision in the branching ratio measurement has been improved from $40\%$ (2016--2018) to $25\%$ (2016--2022).
The NA62 measurements are self-consistent and compatible with the results from the BNL E787 and E949 experiments.
The 2016--2022 NA62 measurement agrees with the SM predictions within $1.7\,\sigma$,
with the central value approximately $50\%$ larger than the SM expectation. 
With more data to be analysed, NA62 aims to reach a $K^{+}\rightarrow\pi^{+}\nu\bar{\nu}$ branching ratio measurement with a relative precision better than $20\%$.

\clearpage
\section*{Acknowledgements}
It is a pleasure to express our appreciation to the staff of the CERN laboratory and the technical
staff of the participating laboratories and universities for their efforts in the operation of the
experiment and data processing.

The cost of the experiment and its auxiliary systems was supported by the funding agencies of 
the Collaboration Institutes. We are particularly indebted to: 
F.R.S.-FNRS (Fonds de la Recherche Scientifique - FNRS), under Grants No. 4.4512.10, 1.B.258.20, Belgium;
CECI (Consortium des Equipements de Calcul Intensif), funded by the Fonds de la Recherche Scientifique de Belgique (F.R.S.-FNRS) under Grant No. 2.5020.11 and by the Walloon Region, Belgium;
NSERC (Natural Sciences and Engineering Research Council), funding SAPPJ-2018-0017,  Canada;
MEYS (Ministry of Education, Youth and Sports) funding LM 2018104, Czech Republic;
BMBF (Bundesministerium f\"{u}r Bildung und Forschung), Germany;
INFN  (Istituto Nazionale di Fisica Nucleare),  Italy;
MIUR (Ministero dell'Istruzione, dell'Universit\`a e della Ricerca),  Italy;
CONACyT  (Consejo Nacional de Ciencia y Tecnolog\'{i}a),  Mexico;
IFA (Institute of Atomic Physics) Romanian 
CERN-RO Nr. 06/03.01.2022
and Nucleus Programme PN 19 06 01 04,  Romania;
MESRS  (Ministry of Education, Science, Research and Sport), Slovakia; 
CERN (European Organization for Nuclear Research), Switzerland; 
STFC (Science and Technology Facilities Council), United Kingdom;
NSF (National Science Foundation) Award Numbers 1506088 and 1806430,  U.S.A.;
ERC (European Research Council)  ``UniversaLepto'' advanced grant 268062, ``KaonLepton'' starting grant 336581, Europe.

Individuals have received support from:
Charles University (grants UNCE 24/SCI/016, PRIMUS 23/SCI/025), 
Ministry of Education, Youth and Sports (project FORTE CZ.02.01.01 \\
/00/22-008/0004632), Czech Republic;
Czech Science Foundation (grant 23-06770S);   
Agence Nationale de la Recherche (grant ANR-19-CE31-0009), France;
Ministero dell'Istruzione, \\
dell'Universit\`a e della Ricerca (MIUR  ``Futuro in ricerca 2012''  grant RBFR12JF2Z, Project GAP), Italy;
the Royal Society  (grants UF100308, UF0758946), United Kingdom;
STFC (Rutherford fellowships ST/J00412X/1, ST/M005798/1), United Kingdom;
ERC (grants 268062,  336581 and  starting grant 802836 ``AxScale'');
EU Horizon 2020 (Marie Sk\l{}odowska-Curie grants 701386, 754496, 842407, 893101, 101023808).


\newpage
\clearpage
\bibliography{bibliography}
\newpage
\clearpage

\newcommand{\orcimg}{\raisebox{-0.3\height}{\includegraphics[height=\fontcharht\font`A]{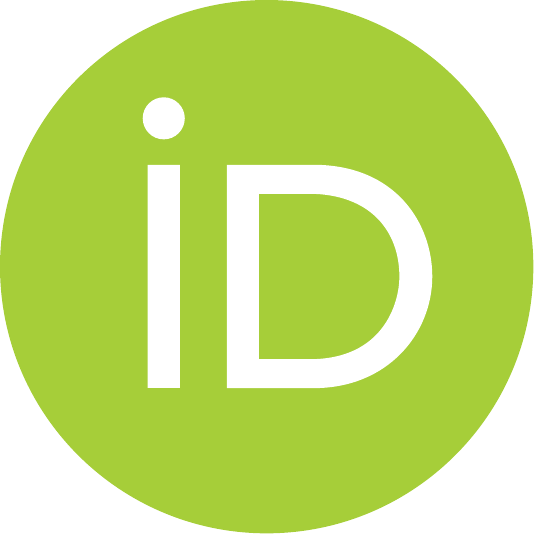}}}
\newcommand{\orcid}[1]{\href{https://orcid.org/#1}{\orcimg}}

\centerline{\bf The NA62 Collaboration} 
\vspace{1.5cm}
%
%

\begin{raggedright}
\noindent
{\bf Universit\'e Catholique de Louvain, Louvain-La-Neuve, Belgium}\\
 E.~Cortina Gil\orcid{0000-0001-9627-699X},
 J.~Jerhot$\,${\footnotemark[1]}\orcid{0000-0002-3236-1471},
 N.~Lurkin\orcid{0000-0002-9440-5927}
\vspace{0.5cm}

{\bf TRIUMF, Vancouver, British Columbia, Canada}\\
 T.~Numao\orcid{0000-0001-5232-6190},
 B.~Velghe\orcid{0000-0002-0797-8381},
 V. W. S.~Wong\orcid{0000-0001-5975-8164}
\vspace{0.5cm}

{\bf University of British Columbia, Vancouver, British Columbia, Canada}\\
 D.~Bryman$\,${\footnotemark[2]}\orcid{0000-0002-9691-0775}
\vspace{0.5cm}

{\bf Charles University, Prague, Czech Republic}\\
 Z.~Hives\orcid{0000-0002-5025-993X},
 T.~Husek$\,${\footnotemark[3]}\orcid{0000-0002-7208-9150},
 K.~Kampf\orcid{0000-0003-1096-667X},
 M.~Kolesar\orcid{0000-0002-9085-2252},
 M.~Koval\orcid{0000-0002-6027-317X}
\vspace{0.5cm}

{\bf Aix Marseille University, CNRS/IN2P3, CPPM, Marseille, France}\\
 B.~De Martino\orcid{0000-0003-2028-9326},
 M.~Perrin-Terrin\orcid{0000-0002-3568-1956},
 L.~Petit$\,${\footnotemark[4]}\orcid{0009-0000-8079-9710}
\vspace{0.5cm}

{\bf Max-Planck-Institut f\"ur Physik (Werner-Heisenberg-Institut), Garching, Germany}\\
 B.~D\"obrich\orcid{0000-0002-6008-8601},
 S.~Lezki\orcid{0000-0002-6909-774X},
 J.~Schubert$\,${\footnotemark[5]}\orcid{0000-0002-5782-8816}
\vspace{0.5cm}

{\bf Institut f\"ur Physik and PRISMA Cluster of Excellence, Universit\"at Mainz, Mainz, Germany}\\
 A. T.~Akmete\orcid{0000-0002-5580-5477},
 R.~Aliberti$\,${\footnotemark[6]}\orcid{0000-0003-3500-4012},
 M.~Ceoletta$\,${\footnotemark[7]}\orcid{0000-0002-2532-0217},
 L.~Di Lella\orcid{0000-0003-3697-1098},
 N.~Doble\orcid{0000-0002-0174-5608}, 
 L.~Peruzzo\orcid{0000-0002-4752-6160},
 C.~Polivka\orcid{0009-0002-2403-8575},
 S.~Schuchmann\orcid{0000-0002-8088-4226},
 H.~Wahl\orcid{0000-0003-0354-2465},
 R.~Wanke\orcid{0000-0002-3636-360X}
\vspace{0.5cm}

{\bf Dipartimento di Fisica e Scienze della Terra dell'Universit\`a e INFN, Sezione di Ferrara, Ferrara, Italy}\\
 P.~Dalpiaz,
 R.~Negrello\orcid{0009-0008-3396-5550},
 I.~Neri\orcid{0000-0002-9669-1058},
 F.~Petrucci\orcid{0000-0002-7220-6919},
 M.~Soldani\orcid{0000-0003-4902-943X}
\vspace{0.5cm}

{\bf INFN, Sezione di Ferrara, Ferrara, Italy}\\
 L.~Bandiera\orcid{0000-0002-5537-9674},
 N.~Canale\orcid{0000-0003-2262-7077},
 A.~Cotta Ramusino\orcid{0000-0003-1727-2478},
 A.~Gianoli\orcid{0000-0002-2456-8667},
 M.~Romagnoni\orcid{0000-0002-2775-6903},
 A.~Sytov\orcid{0000-0001-8789-2440}
\vspace{0.5cm}

{\bf Dipartimento di Fisica e Astronomia dell'Universit\`a e INFN, Sezione di Firenze, Sesto Fiorentino, Italy}\\
 M.~Lenti\orcid{0000-0002-2765-3955},
 P.~Lo Chiatto\orcid{0000-0002-4177-557X},
 I.~Panichi\orcid{0000-0001-7749-7914},
 G.~Ruggiero\orcid{0000-0001-6605-4739}
\vspace{0.5cm}

{\bf INFN, Sezione di Firenze, Sesto Fiorentino, Italy}\\
 A.~Bizzeti$\,${\footnotemark[8]}\orcid{0000-0001-5729-5530},
 F.~Bucci\orcid{0000-0003-1726-3838}
\vspace{0.5cm}

{\bf Laboratori Nazionali di Frascati, Frascati, Italy}\\
 A.~Antonelli\orcid{0000-0001-7671-7890},
 V.~Kozhuharov$\,${\footnotemark[9]}\orcid{0000-0002-0669-7799},
 G.~Lanfranchi\orcid{0000-0002-9467-8001},
 S.~Martellotti\orcid{0000-0002-4363-7816},
 M.~Moulson\orcid{0000-0002-3951-4389}, 
 T.~Spadaro\orcid{0000-0002-7101-2389},
 G.~Tinti\orcid{0000-0003-1364-844X}
\vspace{0.5cm}

{\bf Dipartimento di Fisica ``Ettore Pancini'' e INFN, Sezione di Napoli, Napoli, Italy}\\
 F.~Ambrosino\orcid{0000-0001-5577-1820},
 M.~D'Errico\orcid{0000-0001-5326-1106},
 R.~Fiorenza$\,$\renewcommand{\thefootnote}{\fnsymbol{footnote}}\footnotemark[1]\renewcommand{\thefootnote}{\arabic{footnote}}$^,$$\,${\footnotemark[10]}\orcid{0000-0003-4965-7073},
 M.~Francesconi\orcid{0000-0002-7029-7634},
 R.~Giordano\orcid{0000-0002-5496-7247}, 
 P.~Massarotti\orcid{0000-0002-9335-9690},
 M.~Mirra\orcid{0000-0002-1190-2961},
 M.~Napolitano\orcid{0000-0003-1074-9552},
 I.~Rosa\orcid{0009-0002-7564-1825},
 G.~Saracino\orcid{0000-0002-0714-5777}
\vspace{0.5cm}

{\bf Dipartimento di Fisica e Geologia dell'Universit\`a e INFN, Sezione di Perugia, Perugia, Italy}\\
 G.~Anzivino\orcid{0000-0002-5967-0952}
\vspace{0.5cm}

{\bf INFN, Sezione di Perugia, Perugia, Italy}\\
 P.~Cenci\orcid{0000-0001-6149-2676},
 V.~Duk\orcid{0000-0001-6440-0087},
 R.~Lollini\orcid{0000-0003-3898-7464},
 P.~Lubrano\orcid{0000-0003-0221-4806},
 M.~Pepe\orcid{0000-0001-5624-4010},
 M.~Piccini\orcid{0000-0001-8659-4409}
\vspace{0.5cm}

{\bf Dipartimento di Fisica dell'Universit\`a e INFN, Sezione di Pisa, Pisa, Italy}\\
 F.~Costantini\orcid{0000-0002-2974-0067},
 M.~Giorgi\orcid{0000-0001-9571-6260},
 S.~Giudici\orcid{0000-0003-3423-7981},
 G.~Lamanna\orcid{0000-0001-7452-8498},
 E.~Lari\orcid{0000-0003-3303-0524}, 
 E.~Pedreschi\orcid{0000-0001-7631-3933},
 J.~Pinzino\orcid{0000-0002-7418-0636},
 M.~Sozzi\orcid{0000-0002-2923-1465}
\vspace{0.5cm}

{\bf INFN, Sezione di Pisa, Pisa, Italy}\\
 R.~Fantechi\orcid{0000-0002-6243-5726},
 F.~Spinella\orcid{0000-0002-9607-7920}
\vspace{0.5cm}

{\bf Scuola Normale Superiore e INFN, Sezione di Pisa, Pisa, Italy}\\
 I.~Mannelli\orcid{0000-0003-0445-7422}
\vspace{0.5cm}

{\bf Dipartimento di Fisica, Sapienza Universit\`a di Roma e INFN, Sezione di Roma I, Roma, Italy}\\
 M.~Raggi\orcid{0000-0002-7448-9481}
\vspace{0.5cm}

{\bf INFN, Sezione di Roma I, Roma, Italy}\\
 A.~Biagioni\orcid{0000-0001-5820-1209},
 P.~Cretaro\orcid{0000-0002-2229-149X},
 O.~Frezza\orcid{0000-0001-8277-1877},
 A.~Lonardo\orcid{0000-0002-5909-6508},
 M.~Turisini\orcid{0000-0002-5422-1891},
 P.~Vicini\orcid{0000-0002-4379-4563}
\vspace{0.5cm}

{\bf INFN, Sezione di Roma Tor Vergata, Roma, Italy}\\
 R.~Ammendola\orcid{0000-0003-4501-3289},
 V.~Bonaiuto$\,${\footnotemark[11]}\orcid{0000-0002-2328-4793},
 A.~Fucci,
 A.~Salamon\orcid{0000-0002-8438-8983},
 F.~Sargeni$\,${\footnotemark[12]}\orcid{0000-0002-0131-236X}
\vspace{0.5cm}

{\bf Dipartimento di Fisica dell'Universit\`a e INFN, Sezione di Torino, Torino, Italy}\\
 R.~Arcidiacono$\,${\footnotemark[13]}\orcid{0000-0001-5904-142X},
 B.~Bloch-Devaux$\,${\footnotemark[3]}$^,$$\,${\footnotemark[14]}\orcid{0000-0002-2463-1232},
 E.~Menichetti\orcid{0000-0001-7143-8200},
 E.~Migliore\orcid{0000-0002-2271-5192}
\vspace{0.5cm}

{\bf INFN, Sezione di Torino, Torino, Italy}\\
 C.~Biino$\,${\footnotemark[15]}\orcid{0000-0002-1397-7246},
 A.~Filippi\orcid{0000-0003-4715-8748},
 F.~Marchetto\orcid{0000-0002-5623-8494},
 D.~Soldi\orcid{0000-0001-9059-4831}
\vspace{0.5cm}

{\bf Institute of Nuclear Physics, Almaty, Kazakhstan}\\
 Y.~Mukhamejanov\orcid{0000-0002-9064-6061},
 A.~Mukhamejanova$\,${\footnotemark[16]}\orcid{0009-0004-4799-9066},
 N.~Saduyev\orcid{0000-0002-5144-0677},
 S.~Sakhiyev\orcid{0000-0002-9014-9487}
\vspace{0.5cm}

{\bf Instituto de F\'isica, Universidad Aut\'onoma de San Luis Potos\'i, San Luis Potos\'i, Mexico}\\
 A.~Briano~Olvera\orcid{0000-0001-6121-3905},
 J.~Engelfried\orcid{0000-0001-5478-0602},
 N.~Estrada-Tristan$\,${\footnotemark[17]}\orcid{0000-0003-2977-9380},
 R.~Piandani\orcid{0000-0003-2226-8924},
 M.~A.~Reyes~Santos$\,${\footnotemark[17]}\orcid{0000-0003-1347-2579},
 K.~A.~Rodriguez~Rivera\orcid{0000-0001-5723-9176}
\vspace{0.5cm}

{\bf Horia Hulubei National Institute for R\&D in Physics and Nuclear Engineering, Bucharest-Magurele, Romania}\\
 P.~Boboc\orcid{0000-0001-5532-4887},
 A. M.~Bragadireanu,
 S. A.~Ghinescu\orcid{0000-0003-3716-9857},
 O. E.~Hutanu
\vspace{0.5cm}

{\bf Faculty of Mathematics, Physics and Informatics, Comenius University, Bratislava, Slovakia}\\
 T.~Blazek\orcid{0000-0002-2645-0283},
 V.~Cerny\orcid{0000-0003-1998-3441},
 T.~Velas\orcid{0009-0004-0061-1968},
 R.~Volpe$\,${\footnotemark[18]}\orcid{0000-0003-1782-2978}
\vspace{0.5cm}

{\bf CERN, European Organization for Nuclear Research, Geneva, Switzerland}\\
 J.~Bernhard\orcid{0000-0001-9256-971X},
 L.~Bician$\,${\footnotemark[19]}\orcid{0000-0001-9318-0116},
 M.~Boretto\orcid{0000-0001-5012-4480},
 F.~Brizioli$\,$\renewcommand{\thefootnote}{\fnsymbol{footnote}}\footnotemark[1]\renewcommand{\thefootnote}{\arabic{footnote}}$^,$$\,${\footnotemark[20]}\orcid{0000-0002-2047-441X},
 A.~Ceccucci\orcid{0000-0002-9506-866X}, 
 M.~Corvino\orcid{0000-0002-2401-412X},
 H.~Danielsson\orcid{0000-0002-1016-5576},
 F.~Duval,
 L.~Federici\orcid{0000-0002-3401-9522},
 E.~Gamberini\orcid{0000-0002-6040-4985}, 
 R.~Guida\orcid{0000-0001-8413-9672},
 E.~B.~Holzer\orcid{0000-0003-2622-6844},
 B.~Jenninger,
 Z.~Kucerova\orcid{0000-0001-8906-3902},
 G.~Lehmann Miotto\orcid{0000-0001-9045-7853}, 
 P.~Lichard\orcid{0000-0003-2223-9373},
 K.~Massri$\,${\footnotemark[21]}\orcid{0000-0001-7533-6295},
 E.~Minucci$\,${\footnotemark[22]}\orcid{0000-0002-3972-6824},
 M.~Noy,
 V.~Ryjov, 
 J.~Swallow$\,$\renewcommand{\thefootnote}{\fnsymbol{footnote}}\footnotemark[1]\renewcommand{\thefootnote}{\arabic{footnote}}$^,$$\,${\footnotemark[23]}\orcid{0000-0002-1521-0911},
 M.~Zamkovsky\orcid{0000-0002-5067-4789}
\vspace{0.5cm}

{\bf Ecole Polytechnique F\'ed\'erale Lausanne, Lausanne, Switzerland}\\
 X.~Chang\orcid{0000-0002-8792-928X},
 A.~Kleimenova\orcid{0000-0002-9129-4985},
 R.~Marchevski\orcid{0000-0003-3410-0918}
\vspace{0.5cm}

{\bf School of Physics and Astronomy, University of Birmingham, Birmingham, United Kingdom}\\
 J. R.~Fry\orcid{0000-0002-3680-361X},
 F.~Gonnella\orcid{0000-0003-0885-1654},
 E.~Goudzovski\orcid{0000-0001-9398-4237},
 J.~Henshaw\orcid{0000-0001-7059-421X},
 C.~Kenworthy\orcid{0009-0002-8815-0048}, 
 C.~Lazzeroni\orcid{0000-0003-4074-4787},
 C.~Parkinson\orcid{0000-0003-0344-7361},
 A.~Romano\orcid{0000-0003-1779-9122},
 C.~Sam\orcid{0009-0005-3802-5777},
 J.~Sanders\orcid{0000-0003-1014-094X}, 
 A.~Sergi$\,${\footnotemark[24]}\orcid{0000-0001-9495-6115},
 A.~Shaikhiev$\,${\footnotemark[21]}\orcid{0000-0003-2921-8743},
 A.~Tomczak\orcid{0000-0001-5635-3567}
\vspace{0.5cm}

{\bf School of Physics, University of Bristol, Bristol, United Kingdom}\\
 H.~Heath\orcid{0000-0001-6576-9740}
\vspace{0.5cm}

{\bf School of Physics and Astronomy, University of Glasgow, Glasgow, United Kingdom}\\
 D.~Britton\orcid{0000-0001-9998-4342},
 A.~Norton\orcid{0000-0001-5959-5879},
 D.~Protopopescu\orcid{0000-0002-8047-6513}
\vspace{0.5cm}

{\bf Physics Department, University of Lancaster, Lancaster, United Kingdom}\\
 J. B.~Dainton,
 L.~Gatignon\orcid{0000-0001-6439-2945},
 R. W. L.~Jones\orcid{0000-0002-6427-3513}
\vspace{0.5cm}

\vspace{0.5cm}
{\bf Physics and Astronomy Department, George Mason University, Fairfax, Virginia, USA}\\
 P.~Cooper,
 D.~Coward$\,${\footnotemark[25]}\orcid{0000-0001-7588-1779},
 P.~Rubin\orcid{0000-0001-6678-4985}
\vspace{0.5cm}

{\bf Authors affiliated with an international laboratory covered by a cooperation agreement with CERN}\\
 A.~Baeva,
 D.~Baigarashev$\,${\footnotemark[26]}\orcid{0000-0001-6101-317X},
 V.~Bautin\orcid{0000-0002-5283-6059},
 D.~Emelyanov,
 T.~Enik\orcid{0000-0002-2761-9730}, 
 V.~Falaleev$\,${\footnotemark[18]}\orcid{0000-0003-3150-2196},
 V.~Kekelidze\orcid{0000-0001-8122-5065},
 D.~Kereibay,
 A.~Korotkova,
 L.~Litov$\,${\footnotemark[9]}\orcid{0000-0002-8511-6883}, 
 D.~Madigozhin\orcid{0000-0001-8524-3455},
 M.~Misheva$\,${\footnotemark[27]},
 N.~Molokanova,
 I.~Polenkevich,
 Yu.~Potrebenikov\orcid{0000-0003-1437-4129}, 
 K.~Salamatin\orcid{0000-0001-6287-8685},
 S.~Shkarovskiy
\vspace{0.5cm}

{\bf Authors affiliated with an Institute formerly covered by a cooperation agreement with CERN}\\
 S.~Fedotov,
 K.~Gorshanov\orcid{0000-0001-7912-5962},
 E.~Gushchin\orcid{0000-0001-8857-1665},
 S.~Kholodenko$\,${\footnotemark[28]}\orcid{0000-0002-0260-6570},
 A.~Khotyantsev, 
 Y.~Kudenko\orcid{0000-0003-3204-9426},
 V.~Kurochka,
 V.~Kurshetsov\orcid{0000-0003-0174-7336},
 A.~Mefodev,
 V.~Obraztsov\orcid{0000-0002-0994-3641}, 
 A.~Okhotnikov\orcid{0000-0003-1404-3522},
 A.~Sadovskiy\orcid{0000-0002-4448-6845},
 V.~Sugonyaev\orcid{0000-0003-4449-9993},
 O.~Yushchenko\orcid{0000-0003-4236-5115}
\vspace{0.5cm}

\end{raggedright}

%
%

\setcounter{footnote}{0}
\newlength{\basefootnotesep}
\setlength{\basefootnotesep}{\footnotesep}

\renewcommand{\thefootnote}{\fnsymbol{footnote}}
\noindent
$^{\footnotemark[1]}${Corresponding authors: F.~Brizioli, R.~Fiorenza, J.~Swallow, \\
email: francesco.brizioli@cern.ch, renato.fiorenza@cern.ch, joel.christopher.swallow@cern.ch}\\
\renewcommand{\thefootnote}{\arabic{footnote}}
$^{1}${Present address: Max-Planck-Institut f\"ur Physik (Werner-Heisenberg-Institut), D-85748 Garching, Germany} \\
$^{2}${Also at TRIUMF, Vancouver, British Columbia, V6T 2A3, Canada} \\
$^{3}${Also at School of Physics and Astronomy, University of Birmingham, Birmingham, B15 2TT, UK} \\
$^{4}${Also at Universit\'e de Toulon, Aix Marseille University, CNRS, IM2NP, F-83957 La Garde, France} \\
$^{5}${Also at Department of Physics, Technical University of Munich, M\"unchen, D-80333, Germany} \\
$^{6}${Present address: Institut f\"ur Kernphysik and Helmholtz Institute Mainz, Universit\"at Mainz, Mainz, D-55099, Germany} \\
$^{7}${Also at CERN, European Organization for Nuclear Research, CH-1211 Geneva 23, Switzerland} \\
$^{8}${Also at Dipartimento di Scienze Fisiche, Informatiche e Matematiche, Universit\`a di Modena e Reggio Emilia, I-41125 Modena, Italy} \\
$^{9}${Also at Faculty of Physics, University of Sofia, BG-1164 Sofia, Bulgaria} \\
$^{10}${Present address: Scuola Superiore Meridionale e INFN, Sezione di Napoli, I-80138 Napoli, Italy} \\
$^{11}${Also at Department of Industrial Engineering, University of Roma Tor Vergata, I-00173 Roma, Italy} \\
$^{12}${Also at Department of Electronic Engineering, University of Roma Tor Vergata, I-00173 Roma, Italy} \\
$^{13}${Also at Universit\`a degli Studi del Piemonte Orientale, I-13100 Vercelli, Italy} \\
$^{14}${Present address: Universit\'e Catholique de Louvain, B-1348 Louvain-La-Neuve, Belgium} \\
$^{15}${Also at Gran Sasso Science Institute, I-67100 L'Aquila,  Italy} \\
$^{16}${Also at al-Farabi Kazakh National University, 050040 Almaty, Kazakhstan} \\
$^{17}${Also at Universidad de Guanajuato, 36000 Guanajuato, Mexico} \\
$^{18}${Present address: INFN, Sezione di Perugia, I-06100 Perugia, Italy} \\
$^{19}${Present address: Charles University, 116 36 Prague 1, Czech Republic} \\
$^{20}${Also at INFN, Sezione di Perugia, I-06100 Perugia, Italy} \\
$^{21}${Present address: Physics Department, University of Lancaster, Lancaster, LA1 4YB, UK} \\
$^{22}${Present address: Syracuse University, Syracuse, NY 13244, USA} \\
$^{23}${Present address: Laboratori Nazionali di Frascati, I-00044 Frascati, Italy} \\
$^{24}${Present address: Dipartimento di Fisica dell'Universit\`a e INFN, Sezione di Genova, I-16146 Genova, Italy} \\
$^{25}${Also at SLAC National Accelerator Laboratory, Stanford University, Menlo Park, CA 94025, USA} \\
$^{26}${Also at L. N. Gumilyov Eurasian National University, 010000 Nur-Sultan, Kazakhstan} \\
$^{27}${Present address: Institute of Nuclear Research and Nuclear Energy of Bulgarian Academy of Science (INRNE-BAS), BG-1784 Sofia, Bulgaria} \\
$^{28}${Present address: INFN, Sezione di Pisa, I-56100 Pisa, Italy} \\

\end{document}